\documentclass[12pt]{article}
\usepackage{amssymb}
\evensidemargin \oddsidemargin
\usepackage{graphicx}
\usepackage{amssymb}
\def\nn{\nonumber \\}
\def \T {{\cal T}}
\def \B {{\cal B}}

\def\uo {u_{\scriptscriptstyle 0}}
\def\uu {u_{\scriptscriptstyle 1}}
\def\ho {h_{\scriptscriptstyle 0}}
\def\hp {h_{\scriptscriptstyle \pi}}
\def\ko {k_{\scriptscriptstyle 0}}
\def\kp {k_{\scriptscriptstyle \pi}}

\newcommand{\beq}{\begin{equation}}
\newcommand{\eeq}{\end{equation}}
\newcommand{\bea}{\begin{eqnarray}}
\newcommand{\eea}{\end{eqnarray}}
\newcommand{\ba}{\begin{array}}
\newcommand{\ea}{\end{array}}
\newtheorem{prop}{Proposition}[section]
\newtheorem{lemma}{Lemma}[section]
\newtheorem{theorem}{Theorem}[section]
\def\sq{\mbox{\rlap{$\sqcap$}$\sqcup$}}
\newcommand{\bp}{\begin{proof}}
\newcommand{\ep}{\end{proof}\par\vspace{10pt}\noindent}
\begin{document}

\title{Existence and uniqueness of 
solutions of a class of 3$^{rd}$ order dissipative problems with various 
boundary conditions describing the Josephson effect}
 \author{ Monica De Angelis$^1$,   \hspace{4mm} Gaetano Fiore$^{1,2}$  \\\\
$^{1}$ Dip. di Matematica e Applicazioni, Universit\`a ``Federico II''\\
   V. Claudio 21, 80125 Napoli, Italy\\         %\and
$^{2}$         I.N.F.N., Sez. di Napoli,
        Complesso MSA, V. Cintia, 80126 Napoli, Italy}

\date{}
\maketitle

\begin{abstract}
We prove existence and uniqueness  of solutions of a large class of 
initial-boundary-value problems characterized by a quasi-linear
third order equation (the third order term being dissipative) 
on a finite space interval with Dirichlet, Neumann or pseudoperiodic
boundary conditions.
The class includes equations arising in superconductor theory, such as a 
well-known modified sine-Gordon equation describing the Josephson
effect, and in the theory of viscoelastic materials.  
\end{abstract}

\section{Introduction}
\label{intro}

In this paper we study the class of third order problems
\beq
\ba{l}
 Lu= f(x,t,U), \qquad
L:=\partial_t^2\!+\! a  \partial_t\!-\!c^2\partial_x^2
[\varepsilon \partial_t\!+\!1] , \\[8pt]
u(x,0)= \uo(x), \qquad u_t(x,0)= \uu(x)\ea \quad x\!\in\! \mathring{D},\:\:
t\!>\!0    \label{eqh}
 \eeq
[we have abbreviated \ $U\!:=\!(u,u_x,u_t)$]  \ where the domain $D$ \ 
and the boundary conditions are respectively given by 
\beq
\ba{lll}                                                  \label{23}
D=[0,{\pi}],\quad\:\: &  u(0,t)=\ho(t),\quad\:  u({\pi},t)=\hp(t),
\qquad  &\mbox{DBC},\\[6pt]
D=[0,{\pi}],\quad & u_x(0,t)=\ko(t),\quad\:   u_x({\pi},t)=\kp(t),\qquad
&\mbox{NBC},\\[6pt]
D=\mathbb{R},\qquad & u(x+2\pi ,t)=u(x,t)+ 2\pi m,\qquad  &\mbox{PBC}.
\ea
\eeq
Here and in the sequel: we use DBC, NBC and PBC as abbreviations for
Dirichlet, Neumann and (pseudo)periodic (with $m\!\in\!\mathbb{Z}$) boundary 
conditions respectively; \  $ a,\varepsilon $ are respectively a real and a positive constant;   
$ f$ is continuous and in the PBC case fulfills the compatibility condition
\beq
f(x\!+\!2\pi ,t,u\!+\! 2\pi m,u_x,u_t)=f(x,t,u,u_x,u_t);       \label{mperiodic'}
\eeq
having  set $I\!:=\![0,\infty[$,  $\uo,\uu\in C^2(D)$ 
[and fulfill (\ref{23})$_3$ in the PBC case], \  $\ho,\hp,\in C^2(I)$, 
\ [resp.  \ $\ko,\kp,\in C^1(I)$]  
 \ are assigned so as to be of period $2\pi$ in the PBC case, otherwise so as to
fulfill the consistency matching conditions
\beq
\ba{lllll}
\ho(0)\!=\!\uo(0), \quad & \dot \ho(0)\!=\!\uu(0), \quad &  \hp(0)\!=\!\uo(\!\pi\!), \quad
 & \dot \hp(0)\!=\!\uu(\!\pi\!)\qquad &\mbox{DBC},\\[6pt]
\ko(0)\!=\!\uo'(0), \quad & \dot \ko(0)\!=\!\uu'(0), \quad &  \kp(0)\!=\!\uo'(\!\pi\!),
\quad  & \dot \kp(0) \!=\!\uu'(\!\pi\!)\qquad &\mbox{NBC}.
\ea                          \label{23bis}
\eeq
The  $\varepsilon$-term is dissipative; if $a>0$ the $a$-term is dissipative as well.

Theorems of existence and uniqueness of solutions of  various versions of problem (\ref{eqh})
on $D\!=\!\mathbb{R}$ with $u$ going to zero as $x\!\to\!\pm\infty$ were given in 
\cite{Ren83,DacRen92,DeaRen08,DeaMaiMaz08}. 
Theorems of existence and uniqueness  for some version of problem 
(\ref{eqh}) with DBC (\ref{23})$_1$, as well as for the qualitative properties 
(boundedness, stability, attractivity, 
...) of the solutions, have been given in \cite{DacDan98,Dea01,DanFio01,DanFio05}. 

In this work we prove general results concerning the existence and uniqueness for 
all positive $t$  of the solution  of (\ref{eqh}) with PBC, DBC or NBC . 
In the proof we show also rather stringent properties of the fundamental solutions 
of the equation $Lu=0$ resp. fulfilling the PBC, DBC, NBC.

\begin{figure}[ht]
\begin{center}
\includegraphics[width=6cm]{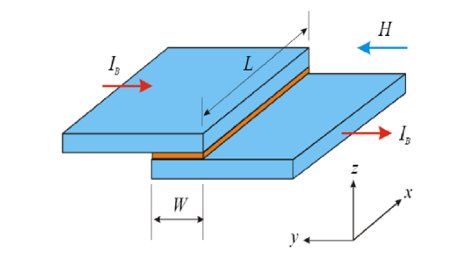}\hskip5cm
\includegraphics[width=3cm]{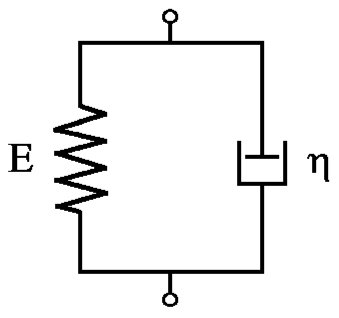}
\end{center}
\caption{Josephson Junction (left) and schematic representation of a Voigt material (right)}
\label{models}       % Give a unique label
\end{figure}

Physically remarkable examples of problems (\ref{eqh}-\ref{23}) include:
\begin{itemize}

\item If $ f\!=\!b\sin u\!-\!\gamma$, with $b,\gamma\!=\!\mbox{const}$,
a modified sine-Gordon  eq. describing
{\bf Josephson effect}  \cite{Jos} in the theory of superconductors, which
is at the base (see e.g. \cite{BarPat82}) of a large number
of advanced developments  both in fundamental research (e.g.
macroscopic effects of quantum physics, quantum computation) and in
applications to electronic devices (see e.g. Chapters 3-6 in
\cite{ChrScoSoe99}): $u(x,t)$ is the phase difference of the
macroscopic wavefunctions of the Bose-Einstein condensate of Cooper  pairs 
in two  superconductors separated by a   {\it Josephson  
junction} (JJ), i.e. a very thin and  narrow   dielectric strip 
of finite length (Fig. \ref{models}-left),
the term  $ a  u_t$ is due to the Joule effect of the  residual  
current of single electrons across the JJ, the term
$\varepsilon u_{xxt}$ is due to the surface impedence of the JJ.
In the simplest  model adopted to describe the  
JJ \  $ a =0$  \ and \ $\varepsilon,c^2\!=\!\mbox{const}$
($\varepsilon$ is rather small);   more   accurately, 
$ a$ is positive but very small; even  more  accurately, 
one adopts $f\!=\!b\sin u\!-\!\gamma\!+\!a(1\!-\!\cos u)u_t$.
The equation must be complemented by NBC if the strip is open, 
by PBC if the strip is closed in the form of a ring.
In the latter case the topological invariant $m$ counts the number of {\it fluxons}
(i.e. quanta of magnetic flux) which may move along the ring but remain trapped in it,
as the closed lines of the magnetic field passing through the junction and encircling
one of the two  superconductors cannot escape crossing it, by the Meissner effect.

\item If \ $ a \!=\!0$, \ $f=f(x,t)$, \ an equation (see e.g.
\cite{Mor56,Ren83}) for the displacement $u(x,t)$ 
of the section of a rod from its rest position $x$ 
in a Voigt material: \ $f$ is applied density force, 
$c^2\!=\!E/\rho$, $\varepsilon\!=\!1/\rho\mu$, where $\rho$ is the linear 
density of the rod at rest, $E,\mu$ are the elastic and
viscous constants of the rod, which enter the stress-strain relation
$\sigma=E\nu+\partial_t \nu/\mu$, where $\sigma$ is the stress
$\nu$ is the strain 
(as known, a discretized model of the rod is a series
of elements consisting of a viscous damper and an elastic spring connected in parallel 
as shown in Fig. \ref{models}-right).

\item  Equations used to describe: heat conduction at low temperature $u$ 
\cite{MorPayStr90,FlaRio96}, if  \ $\varepsilon \!=\!c^2$,  $f=0$; sound
propagation in viscous gases \cite{Lam32}; propagation of plane
waves in perfect incompressible and electrically conducting fluids
\cite{Nar53}.

\end{itemize}

We may assume without loss of generality
that $c^2\!=\!1$ [this can be always obtained by the
rescaling (\ref{redef0}) with $a=0$ of $x$].  Redefining
\beq
\ba{ll}                                                  \label{redef}
\hat u(x,t)=u(x,t)-\phi_m(x),\qquad  \qquad\:\: &\mbox{PBC},\\[6pt]
\hat u(x,t)=u(x,t) +\left[\frac {x^2}{2\pi}-x\right]\ko(t)
- \frac {x^2}{2\pi} \kp(t),\quad \qquad\:\:
&\mbox{NBC},\\[6pt]
\hat u(x,t)=u(x,t) +\left[\frac x\pi-1\right]\ho(t)- \frac x\pi
\hp(t),\qquad \qquad\:\:  &\mbox{DBC},
\ea
\eeq
where $\phi_m(x)$ is such that
$\phi_m(x\!+\!2\pi)=\phi_m(x)\!+\!2\pi m$, e.g. $\phi_m(x):=mx$,
we find that $\hat u$ fulfills the PDE, initial conditions
\beq                                                   \label{22'}
L \hat u=\hat  f,\qquad \qquad\hat u(x,0)=\hat \uo(x),
\qquad \qquad \hat u_t(x,0)=\hat \uu(x)
\eeq
and the boundary conditions
\beq
\ba{lll}                                                  \label{23'}
\hat u(x\!+\!2\pi ,t)=\hat u(x,t),\qquad  \qquad\:\: &&\mbox{PBC},\\[6pt]
\hat u(0,t)=0,\qquad & \hat u({\pi},t)=0,\qquad\:\:  &\mbox{DBC},\\[6pt]
\hat u_x(0,t)=0,\qquad &  \hat u_x({\pi},t)=0,\qquad\:\:
&\mbox{NBC},
\ea
\eeq
with \ $\hat f$ \ and initial conditions respectively given by
\beq
\ba{lll}                                                  \label{redef'}
\hat f(x,\!t,\!\hat u,\!\hat u_x,\!\hat u_t)=f\big[x,\!t,\!\hat u\!+\!\phi^m\!(\!x\!),\!\hat u_x,\!\hat u_t\big]
\!+\!\phi^m_{xx}\!(\!x\!),
\quad & \hat \uo=\uo\!-\!\phi^m(x),
\quad \hat \uu=\uu \quad\:\: &\mbox{PBC},\\[16pt]
 \hat f \!= \!f \!\!+\!\!\left[ \!\frac x\pi\!\!-\!\!1 \!\right]\!\!(\!\ddot \ho
\!+\! a \dot \ho\!)\!-\! \frac x\pi(\!\ddot \hp\!\!+\! a \dot \hp\!)
,\quad & \left\{\!\! \ba{l}
\hat \uo\!=\!\uo \!+\!\!\left[\!\frac x\pi\!\!-\!\!1\!\right]
\!{\scriptstyle \ho(0)}\!-\! \frac x\pi {\scriptstyle \hp(0)}\\[4pt]
\hat \uu\!=\!\uu \!+\!\!\left[\!\frac x\pi\!\!-\!\!1\!\right]
\!{\scriptstyle \dot \ho(0)}\!-\! \frac x\pi{\scriptstyle \dot \hp(0)}
\ea\right. \quad  &\mbox{DBC},\\[16pt]
\hat f \!= \!f \!\!+\!\! \left[ \!\frac {x^2}{2\pi}\!-\!x \!\right]\!\!(\!\ddot \ko
\!+\! a \dot \ko\!)\!-\! \frac {x^2}{2\pi}(\!\ddot \kp\!+\! a \dot \kp\!)
,\quad & \left\{\!\! \ba{l}
\hat \uo\!=\!\uo \!+\!\left[\!\frac {x^2}{2\pi}\!-\!x\!\right]\! {\scriptstyle \ko(0)}
\!\!-\! \frac {x^2}{2\pi} {\scriptstyle \kp (0)}\\[6pt]
\hat \uu=\uu \!+\!\left[\!\frac {x^2}{2\pi}\!-\!x\!\right]\!{\scriptstyle \dot \ko(0)}
\!\!-\! \frac {x^2}{2\pi}{\scriptstyle  \dot \kp(0)}\ea\right.
\quad &\mbox{NBC}.
\ea
\eeq
 $\hat \uo,\hat \uu$ automatically fulfill the consistency conditions
\beq
\ba{lllll}
\hat \uo(0)\!=\!0, \quad & \hat \uu(0)\!=\!0, \quad &  \hat \uo(\!\pi\!)\!=\!0, \quad
 & \hat \uu(\pi)\!=\!0\qquad &\mbox{DBC},\\[6pt]
\hat \uo'(0), \quad & \hat \uu'(0)\!=\!0, \quad &  \hat \uo'(\!\pi\!)\!=\!0,
\quad  & \hat \uu'(\!\pi\!)\!=\!0\qquad &\mbox{NBC}.
\ea                          \label{23bis'}
\eeq
Consequently, without loss of generality we can assume in (\ref{23}) \ $m\!=\!0$, \ 
$h_i\!\equiv\! 0$ \  and  \ $k_i\!\equiv\! 0$ ($i\!=\!0,{\pi}$) respectively
for the PBC, DBC, NBC, namely assume the boundary conditions  (\ref{23'}), (\ref{23bis'}).
Note that in the PBC case from  (\ref{mperiodic'}) it follows \
\ $\hat f(x\!+\!2\pi ,t,U)=\hat f(x,t,U)$, \ as it must be. 
We shall remove the superscripts \ \ $\hat{}$ \ \ henceforth.

We may also assume without loss of generality that $a\!\ge\!0$, $c\!=\!1$. \  In fact, 
if $a\!<\!0$ then $1\!-\!\frac a2\varepsilon\!>\!0$, so that redefining
\beq
\ba{l}
\tilde c:=c\sqrt{1\!-\!\frac a2\varepsilon},\qquad
\tilde x:= \frac 1{\tilde c} x,\qquad
\tilde \varepsilon:= \frac {\varepsilon}{1\!-\!\frac a2\varepsilon},\qquad
\tilde L:=\partial_t^2\!-\!\partial_{\tilde x}^2
(\tilde\varepsilon \partial_t\!+\!1), \\[10pt]
\tilde u(\tilde x,t):=e^{\frac a2 t}u\left( \tilde c\tilde x,t\right),
\qquad\tilde \uo(x):=\uo\!\left(\tilde c\tilde x\right),\qquad
\tilde \uu(x):=\uu\!\left(\tilde c\tilde x\right),\\[12pt]
\tilde f(\tilde x,t,\tilde u,\tilde u_{\tilde x}, \tilde u_t)
\!:=\!\frac {a^2}4\tilde u \!+\!e^{\frac a2 t}f\!\left[\tilde c\tilde x,t,e^{\frac {-a}2 t}\tilde u, \frac {e^{-\frac a2 t}}{\tilde c}\tilde u_{\tilde x}, 
e^{\frac {-a}2 t}(\tilde u_t\!-\!\frac a2\tilde u)\!\right]\!,\\[12pt]
\ea                                                  \label{redef0}
\eeq
we find that $\tilde u$ fulfills the PDE, initial conditions
\beq                                                   \label{22''}
\tilde L \tilde u=\tilde  f,\qquad \qquad\tilde u(\tilde x,0)=\tilde \uo(\tilde x),
\qquad \qquad \tilde u_t(\tilde x,0)=\tilde \uu(\tilde x),
\eeq
and again boundary conditions of the type (\ref{23'}), (\ref{23bis'}).
We shall remove the superscripts \ \ $\tilde{}$ \ \ henceforth.

\section{The fundamental solutions of $Lu=0$}

By  saying that \ $v(x,t)$ \  is a solution of \ $L v\!=\!0$ \ we mean that  \
$v,v_t,v_{tt},\partial_x^2(\varepsilon v_t\!+\!v)$ \ 
are continuous and the combination $Lv$ is zero for \ $t>0$. \
Any solution $ u^d$ of $Lu=0$ and the DBC (\ref{23'}) [resp. $ u^n$ of  $Lu=0$
and the NBC (\ref{23'})] can be transformed by an odd (resp. even) extension into a solution 
$ u^p$  of $Lu=0$ and the PBC (\ref{23'}), as follows. As a first step,
\beq
\ba{ll} 
 u^p(x,t):=\left\{\ba{lll}   & u^d(x,t)\qquad & x\in[0,\pi]\\[6pt]
 -& u^d(-x,t)\qquad & x\in]-\pi,0[          \ea\right.\qquad\qquad &\mbox{DBC},\\[12pt]
 u^p(x,t):=\left\{\ba{lll}   & u^n(x,t)\qquad & x\in[0,\pi]\\[6pt]
 & u^n(-x,t)\qquad & x\in]-\pi,0[          \ea\right.\qquad\qquad &\mbox{NBC};\ea
                                                        \label{evenoddext}
\eeq
as a second step, in either case setting for  $x\in]-\pi,\pi]$ and any $k\in\mathbb{Z}$
\beq
 u^p(x+2k\pi,t):= u^p(x,t).                           \label{evenoddext'}
\eeq
It is immediate to check that, because of (\ref{23'}), $u^p,u^p_x$ and their 
first and second time derivatives are continuous at all points $x=k\pi$; moreover,
because of $Lu^d=0$ (resp. $Lu^n=0$), then $Lu^p=0$
everywhere and, since $u^p_t,u^p_{tt}$ are continuous also at all points $x=k\pi$,
also  $\partial_x^2(\varepsilon u^p_t\!+\!u^p)$ is continuous, as claimed.
Therefore the fundamental solutions of $Lu=0$ and the DBC, NBC   (\ref{23'}) 
can be extended as particular
solutions of  $Lu=0$ and the PBC  (\ref{23'}). 

\medskip
For all \ $n\!\in\!\mathbb{Z}$ \ the products \ $v_n(x,t)\!=\!H_n(t)e^{inx}$ \ are periodic solutions 
of  \ $L v\!=\!0$ provided
\beq
\ddot H_n+ a \dot H_n+n^2(\varepsilon \dot H_n\!+\!H_n)=0;
%\qquad\qquad H_n(0)=0,\quad\dot H_n(0)=1
 \label{ode}
\eeq
then also $v_{nt}$ are.
We choose the solutions fulfilling the  initial conditions \ $H_n(0)\!=\!0$, \ $\dot H_n(0)\!=\!1$:
\beq                                                                          
H_n(t):=e^{-h_n t}\frac {\sinh (\omega_n t)}{\omega_n},\qquad\quad
h_n=\frac { a +\varepsilon n^2}2,\qquad \omega_n=\sqrt{h_n^2-n^2},           \label{defHn}
\eeq
in particular \ $H_0(t):=\frac{1\!-\!e^{- a  t}}{ a }$;
\  $H_n$ must be understood as
its \ $\omega_n\!\to\! 0$ \ limit when \ $\omega_n\!=\!0$:
\beq
H_n(t):=e^{-h_n t}t\qquad \quad \mbox{if }\omega_n=0. \label{defHn0}
\eeq
Clearly \ $h_{-n}\!=\!h_n$, \  $\omega_{-n}\!=\!\omega_n$, \ $H_{-n}\!=\!H_n$; \
note that  $H_n$ 
% is invariant under the replacement $\omega_n\mapsto -\omega_n$ and 
is real even when $\omega_n$ is imaginary.  Any finite
combination of the $v_n,v_{nt}$ is a periodic solution of  \ $L v\!=\!0$.
 We now inquire if also the following (Fourier) series defines one:
\beq
\vartheta(x,t) :=\frac 1{2\pi}\sum\limits_{n\in\mathbb{Z}} H_n(t)e^{inx}
=\frac 1{2\pi}H_0(t)+\frac 1{\pi}\sum\limits_{n=1}^\infty H_n(t)\cos(n x). \label{defvartheta}
\eeq

\begin{prop}
If $a\ge 0$ the series (\ref{defvartheta}) defines for all \ $t\ge 0$ \ 
a continuous, real-valued function \ $\vartheta(x,t)$,  \ even and of period $2\pi$ w.r.t. $x$, 
such that \ $\vartheta(x,0)\!\equiv\! 0$, \ bounded as follows: 
\bea
&& 2\pi|\vartheta(x,t)|\le 2\pi\vartheta(0,t)\le N(t):= M+\left\{\ba{ll}a^{-1}\quad &\mbox{if } a>0,\\
t\quad &\mbox{if } a=0,  \ea     \right.     \label{thetaineq}\\[10pt]
&& M:=2\!+\! 2\log\bar n \!+\!\frac{2\pi^2}{3\varepsilon},  \qquad  \qquad
\bar n:=1\!+\!\left[\frac 2\varepsilon\right];   \label{defbarn}
\eea
here $[y]$ means the integer part of $y$.
For \ $t\!>\!0$ \ the derivatives $\vartheta_t,\vartheta_{tt},\vartheta_{ttt}$, 
$\partial_x^2(\varepsilon\vartheta_t\!+\!\vartheta)$, 
$\partial_x^2(\varepsilon\vartheta_{tt}\!+\!\vartheta_t)$ are well-defined,
equal the term-by-term derived series and fulfill
\beq
L\vartheta=0,\qquad \qquad L\vartheta_t=0
\label{thetaprop}
\eeq
for all $(x,t)\in\mathbb{R}\times\mathbb{R}^+$.
%For \ $t\!>\!0$ \ $\vartheta_t(x,t)$ \ is bounded as follows,
%\beq
%2\pi |\vartheta_{t}(x,t) | \le 1\!\!+\!\!2\!\left(\!\frac 2{\varepsilon}\!+\!1\!\right)^2\! 
%\!+\!\frac 3{\varepsilon}\!+\! 2e^{\frac 2\varepsilon t}\,
%\theta\!\left(0,i\frac \pi\varepsilon t\right),                         \label{thetatineq}
%\eeq 
Finally,  \ $\vartheta_x(\cdot,t)\!\in\! L^2([0,\pi])$ for all $t\!\ge\! 0$, \ 
$\vartheta_{t}(\cdot,t),\vartheta_{xt}(\cdot,t)\!\in\! L^2([0,\pi])$ for all $t\!>\! 0$, with
square $L^2$-norms bounded as follows:
\bea
&&\ba{l}
4\pi^2\Vert \vartheta_x(\cdot,\!t) \Vert_2^2<
2\!+\! \frac4{\varepsilon}\!+\! \frac{4\pi^2}{3\varepsilon^2}, \\[12pt]
4\pi^2\Vert \vartheta_{t}(\cdot,t) \Vert_2^2
<\kappa \!+\!8e^{\frac {4t}\varepsilon}\,\theta\!\left(0,i\frac 2\pi\varepsilon t\right) \!, \qquad \qquad
\kappa\!:=\!\!3\!+\!\frac 4{\varepsilon}\!+\!\frac {2\pi^2}{9\varepsilon^2},\\[12pt]
4 \pi^2\Vert \vartheta_{tx}(\cdot,t) \Vert_2^2
< \left(\!\frac 2{\varepsilon}\!+\!1\!\right)^4\!\left[\!\left(\!\frac 2{\varepsilon}\!+\!1\!\right)^4\!\!+\!1\right]\!+\!\frac {12}{\varepsilon^2}
-\!\frac 8\varepsilon \, e^{\frac 4\varepsilon t}\,
\partial_t\!\left[\theta\!\left(0,\!i\frac {2\varepsilon}\pi t\right)\right];
\ea            \quad    \label{thetax}
\eea
\label{prop1}
here  \  $\|g\|^2_2\!:=\!\int_0^{2\pi} \!\! \frac {dx}{2\pi}\, |g(x)|^2$, and
\ $\theta(z,\tau)\!:=\!\sum\limits_{n\in\mathbb{Z}}\! e^{i\pi(2nz+n^2\tau)}$ \ is Jacobi Theta function. 
\end{prop}
We recall that for $\eta\!>\!0$ \ $\theta\!\left(0,i\eta\right)>0$ \ and \
$\theta\!\left(0,i\eta\right)\sim \eta^{-\frac 12}$ \ as \ $\eta\!\downarrow\! 0$.

If $a\!<\!0$ one could show that the above bounds hold adding at the rhs
some  term proportional to $e^{-at}$ (which is increasing with $t$), while
all other claims hold unmodified. We don't need to do, 
because by the change of variables (\ref{redef0})  we reduce 
the existence and uniqueness theorem for the case $a<0$ to the one for  the case 
 $a=0$ (see section \ref{exist}).

In order to prove the proposition we first prove

\begin{lemma} 
If $a\ge 0$ $H_n$ and its time derivatives fulfill the following bounds:
\bea
%\label{disegHnl}
&& \ba{l}
\left\vert\frac {d^lH_n}{dt^l}\right\vert  \le  \frac {1}{2\omega_n} 
\left[\frac {2^l}{\varepsilon^l} \!+\! (a \!+\!\varepsilon n^2)^l
e^{-t \left(\varepsilon n^2-\frac 2\varepsilon\right) }\right]  \\[10pt]
n^2\vert \varepsilon\dot H_n \!+\!H_n\vert \le   \frac 1{2\omega_n} 
\left[\frac{a\varepsilon \!+\!4}{\varepsilon^2 n^2}
 %e^{-\frac{4t}{3\varepsilon}} 
+n^2( a\varepsilon \!+\!\varepsilon^2 n^2\!+\!1)\,
e^{-t \left(\varepsilon n^2-\frac 2\varepsilon\right) }\right]\\[10pt]
n^2\vert \varepsilon\ddot H_n \!+\!\dot H_n\vert \le  
 \frac 1{2\omega_n} \left[\frac{8\!+\! 2a \varepsilon}{\varepsilon^3}
 %e^{-\frac{4t}{3\varepsilon}} 
+n^2( a\varepsilon \!+\!\varepsilon^2 n^2\!+\!1)(a \!+\!\varepsilon n^2)\,
e^{-t \left(\varepsilon n^2-\frac 2\varepsilon\right) }\right]  %\label{disegHneps}
\ea  \quad{\rm if }\:\: |n|\!\ge\! \bar n,\qquad \label{disegHnl}\\ [12pt]
&&\qquad \qquad\left| H_n(t)\right| \le \left\{\ba{ll} \frac 1{|n|}\qquad\quad  
&{\rm if }\:\:0<|n|\!\le\! \bar n, \\[10pt]
\frac{2}{\varepsilon} \frac1{n^2}\qquad\quad  
&{\rm if }\:\: |n|\!\ge\! \bar n.\ea \right.         \label{ineqn} \\[8pt]
&& \qquad\qquad\left|H_n(t)\right|\le  t,       \label{Hnineq} \\[8pt]
&& \qquad\qquad\big|\dot H_n(t)\big| \le 1,    %\qquad\qquad\qquad \forall n\in\mathbb{Z},   
\label{genineq'} \\[10pt]
&&\qquad\qquad |1\!-\!\dot H_n(t)|< (2h_n \!+\! |\Im (\omega_n)|)\, t.
 \label{buona'}
\eea
\end{lemma}

\noindent
Clearly the bounds (\ref{Hnineq}), (\ref{buona'}) are stringent for $t\sim 0$.

{\it Proof.} \ \
If $l=0,1,2,....$ and $\omega_n\neq 0$ we find
\bea
&&\frac {d\,{}^l H_n}{dt^l}=\frac {1}{2\omega_n}\left[(\omega_n\!-\!h_n)^l
e^{(\omega_n\!-\!h_n)t}- (-)^l(\omega_n\!+\!h_n)^le^{-(\omega_n\!+\!h_n)t}    
\right],\nn
&&\varepsilon\dot H_n \!+\!H_n= \frac 1{2\omega_n}\left[(1\!+\!
\varepsilon\omega_n\!-\!\varepsilon h_n)
e^{(\omega_n\!-\!h_n)t}+ (\varepsilon \omega_n\!+\!\varepsilon h_n\!-\!1)
e^{-(\omega_n\!+\!h_n)t}\right],\qquad           \label{dHn}\\
&&\varepsilon\ddot H_n \!+\!\dot H_n= \frac 1{2\omega_n}\left[(1\!+\!
\varepsilon\omega_n\!-\!\varepsilon h_n)(\omega_n\!-\!h_n)
e^{(\omega_n\!-\!h_n)t}
-(\varepsilon \omega_n\!+\!\varepsilon 
h_n\!-\!1)(\omega_n\!+\!h_n) e^{-(\omega_n\!+\!h_n)t}\right],\qquad \nonumber
\eea
whence
\bea
&& \left\vert\frac {d^lH_n}{dt^l}\right\vert \le
\frac {1}{2|\omega_n|}\left[|h_n\!-\!\omega_n|^l\left|  e^{(\omega_n\!-\!h_n)t} \right|
+ |\omega_n\!+\!h_n|^l \left| e^{-(\omega_n\!+\!h_n)t} \right| \right],\nn
&& n^2\vert \varepsilon\dot H_n \!+\!H_n\vert \le \frac {n^2}{2|\omega_n|}\left[|1\!+\!
\varepsilon\omega_n\!-\!\varepsilon h_n|
\left| e^{(\omega_n\!-\!h_n)t} \right|+ |\varepsilon \omega_n\!+\!\varepsilon h_n\!-\!1|
\left| e^{-(\omega_n\!+\!h_n)t} \right|\right],\qquad\quad \label{disegHnl0}\\
&& n^2\vert \varepsilon\ddot H_n \!+\!\dot H_n\vert \le \frac {n^2}{2|\omega_n|}\left[|1\!+\!
\varepsilon\omega_n\!\!-\!\varepsilon h_n|\, |h_n\!\!-\!\omega_n|
\left| e^{(\omega_n\!-\!h_n)t} \right|
\right. \nn &&\qquad\left.  \qquad\qquad\quad
+|\varepsilon \omega_n\!\!+\!\varepsilon h_n\!\!-\!1|\, |\omega_n\!+\!h_n|
\left| e^{-(\omega_n\!+\!h_n)t} \right|\right]. \qquad \nonumber
 \eea
We recall that for $0<\sigma<1$
\beq
\qquad\qquad\qquad\left.\ba{r}1\!-\!\sigma\\
 1\!-\!\frac{\sigma}2\!-\!\frac{\sigma^2}2\ea
\right\} <\sqrt{1-\sigma}< 1\!-\!\frac{\sigma}2;
\label{diseq0}
\eeq
these inequalities follow from their squares, and become equalities for $\sigma\!=\!0$.
Clearly, \ $h_n\to\infty$, \ $\frac{n^2}{h_n^2}\to0$ \ as  \ $|n|\to\infty$,  \ hence
there exists a  \ $\bar n\!\in\!\mathbb{N}$  \ such that % \ $|n|\!\ge\! \bar n$ \ implies
\beq
|n|\ge \bar n\qquad\quad\Rightarrow\qquad\quad h_n\!>\!0, \quad\:\frac{n^2}{h_n}\!\le \!
\frac 2{\varepsilon},\quad\: \sigma_n:=\frac{n^2}{h_n^2}\!\in]0,1[, 
\quad \: \frac{\omega_n}{h_n}=\sqrt{1\!-\!\sigma_n}\!>\!0.       \label{diseq1}
\eeq
Note also that for $n\ge \bar n$  the sequences $h_n, \frac{\omega_n}{h_n}$
are increasing with $n$, while the sequence $\sigma_n$ is decreasing.
One can choose $\bar n$ as in (\ref{defbarn}).\footnote{
The first two inequalities (\ref{diseq1}) are automatic.  \  The fourth
relation is a consequence of the third. The latter holds iff $\bar n/h_{\bar n}<1$,
 or equivalently $\varepsilon \bar n^2\!-\!2\bar n\!+\!a\!>\!0$; this is
satisfied by all $\bar n\ge 1$ if $a\varepsilon>1$, because then the solutions 
\beq
n_\pm=(1\!\pm\!\sqrt{1\!-\!a\varepsilon})/\varepsilon   \label{meq}
\eeq
 of the equation \ $\varepsilon m^2\!-\!2m\!+\!a\!=\!0$
are not real; otherwise it is satisfied if we choose $\bar n>n_+$, in particular 
(\ref{defbarn}).
}
From (\ref{diseq0}-\ref{diseq1})  we  find for  $|n|\!\ge\! \bar n$
\bea
&& \left.\ba{r}1\!-\!\frac{n^2}{h_n^2}\\[4pt]
1\!-\!\frac{n^2}{2h_n^2}\!-\!\frac{n^4}{2h_n^4}\ea\right\}
\le  \frac{\omega_n}{h_n} \le 1\!-\!\frac{n^2}{2h_n^2}, \label{diseg1}\\
&& 0\le \frac{n^2}{2h_n}\le h_n -\omega_n \le 
\left\{\ba{l}
\frac{n^2}{h_n} \le\frac 2{\varepsilon}\\[4pt]
\frac{n^2}{2h_n}\!+\!\frac{n^4}{2h_n^3},\ea\right.   \label{diseg2}\\ 
&&\left.\ba{r} 0\\[4pt] a \!+\!\varepsilon n^2\!-\!\frac 2  \varepsilon\ea\right\}\le\omega_n\!+\!h_n\le 
 a \!+\!\varepsilon n^2,\label{diseg3}\\
&& \frac{ a \!-\!\frac{4}{\varepsilon}}{ a \!+\!\varepsilon n^2}
\le \frac{ a \!-\!\frac{n^4\varepsilon}{h_n^2}}{ a \!+\!\varepsilon n^2}
=1\!-\!\frac{n^2\varepsilon}{2h_n}\!-\!\frac{n^4\varepsilon}{2h_n^3}\nn
&& \le 1+\varepsilon(\omega_n\!-\!h_n)
\le 1\!-\!\frac{n^2\varepsilon}{2h_n}=\frac{ a }{ a \!+\!\varepsilon n^2},\nn
&& \Rightarrow \qquad|1+\varepsilon(\omega_n\!-\!h_n)|\le  
\frac{a\varepsilon \!+\!4}{\varepsilon^2 n^2}, \label{diseg4}
%\\&&\frac{h_n}{\omega_n}\!-\!1=\frac 1{\omega_n}(h_n-\omega_n)\le 
%\frac{8}{\varepsilon^2 n^2}.
\eea
Each of the inequalities (\ref{diseg2}-\ref{diseg4}) is based on the preceding ones, (\ref{defHn})
or $\omega_n\!+\!h_n\!=\!\omega_n\!-\!h_n\!+\!2h_n$. 
Formulae (\ref{disegHnl0}-\ref{diseg4}) imply  (\ref{disegHnl}).

We now better evaluate the upper bound on $H^2_n(t)$. Except in the case
$a=n=0$, this is a smooth function vanishing at $t=0$ and going to zero as $t\to \infty$.
Its  maximum is reached at the 
smallest solution\footnote{In fact, there is only one solution for $n$ such that $\omega_n$ is real.}
 \ $t=t_{n}\ge 0$ \   of the eq. \ $\dot H_n(t)\!=0$,  \
i.e. of
$$
\ba{ll}
e^{2\omega_n t_{n}}=\frac{h_n\!+\!\omega_n}{h_n\!-\!\omega_n}
=\frac{(h_n\!+\!\omega_n)^2}{n^2}\qquad\qquad 
%e^{\omega_n t_{n}}=\frac{h_n\!+\!\omega_n}{|n|}
\qquad &\mbox{if } \omega_n\neq 0,\\[10pt] 
t_{n}=\frac 1{h_n}\qquad &\mbox{if } \omega_n= 0
\ea  %\label{maxpoint} 
$$
whence it follows in either case\footnote{In fact, if $\omega_n\neq 0$ we find
\bea
&&\left\vert H_n(t_{n})\right\vert = 
\frac {e^{-h_nt_{n}}|e^{-\omega_n t_{n}}|}{2|\omega_n|} \left\vert
e^{2\omega_n t_{n}}-1 \right\vert
=\frac {e^{-h_nt_{n}}|n|}{2|\omega_n(h_n\!+\!\omega_n)|} \left\vert
\frac {(h_n\!+\!\omega_n)^2}{n^2}\!-\! 1\right\vert \nn
&&=\frac {e^{-h_nt_{n}}|n|^{-1}}{2|\omega_n(h_n\!+\!\omega_n)|} \left\vert
h_n^2\!+\!\omega_n^2\!+\!2h_n\omega_n \!-\! n^2\right\vert= e^{-h_nt_{n}}|n|^{-1},\nonumber
\eea
whereas if $\omega_n= 0$ it is $h_n=|n|$ and again \ \
$\left\vert H_n(t_{n})\right\vert =t_{n}e^{-h_nt_{n}}
=h_n^{-1}e^{-h_nt_n}= e^{-h_nt_{n}}|n|^{-1}$.
}
\beq
\left\vert H_n(t)\right\vert \le \left\vert H_n(t_{n})\right\vert
=e^{-h_nt_{n}}|n|^{-1}.        \label{genineq} 
\eeq
From (\ref{genineq}) and $h_nt_{nl}\ge 0$ it follows (\ref{ineqn}) with  $|n|\!\le\! \bar n$.
For $|n|\!\ge\! \bar n$ \ $\omega_n$ is real and we  find 
\beq
e^{-h_nt_{n}}=\left[e^{-\omega_nt_{n}}\right]^{h_n/\omega_n}
=\left[\frac{|n|}{h_n\!+\!\omega_n}\right]^{h_n/\omega_n}\le \frac{|n|}{h_n},
\eeq
whence (\ref{ineqn}) follows by (\ref{genineq}).
On the other hand,
we recall the inequality\footnote{Set $z=x+iy$. (\ref{utile'}) can be proved 
in three steps: 
\bea
&& x>0\qquad\Rightarrow\qquad e^{-x}<1\qquad\Rightarrow\qquad
1\!-\!e^{-x}=\int^x_0\!\!dx'\, e^{-x'}<\int^x_0\!\!dx'=x\label{bb}\\
&& y>0\qquad\Rightarrow\qquad \sin y<y\qquad\Rightarrow\qquad
0\le 1\!-\!\cos y=\int^y_0\!\!dy'\, \sin y'<\frac {y^2}2\nn[8pt]
&& \left| 1\!-\!e^{-z}\right|^2=(1\!-\!e^{-x-iy})(1\!-\!e^{-x+iy})
=(1\!-\!e^{-x})^2+2e^{-x}(1\!-\!\cos y)<x^2
\!+\!y^2=|z|^2 . \nonumber
\eea
}
\beq
 \left| 1\!-\!e^{-z}\right|\le |z|\qquad \qquad \mbox{if }\:
\Re z\ge 0 \label{utile'}
\eeq
(the inequality is strict iff $z\neq 0$); applying it to relation (\ref{defHn}) 
we find for all $n\in\mathbb{Z}$ \ 
$\left|H_n(t)\right|\le \left|\frac{1\!-\!e^{-2\omega_n t}}{2\omega_n}\right|\le t$,  \ i.e. (\ref{Hnineq}).

We now better evaluate the bounds on $\dot H_n(t)$. 
From the definitions (\ref{defHn}-\ref{defHn0}) it easily follows
\beq
\dot H_n(t)= e^{-h_nt}\cosh(\omega_nt)\!-\!h_nH_n(t)        \label{1dHn} 
\eeq
For  any $n\in\mathbb{Z}$ \  this 
is a bounded function equal to 1 at $t=0$ and going to 0 as $t\to \infty$. 
Its  infimum and supremum are reached either at $0,\infty$ or at the
solutions \ $t=t_n'\ge 0$ \  of the eq. \ $\ddot H_n\vert_{t=t_n'}\!\!=0$,  \
i.e. of
$$
\ba{ll}
e^{2\omega_n t_n'}=\left(\!\frac{h_n\!+\!\omega_n}{h_n\!-\!\omega_n}\!\right)^{2}\!\!
=\left[\frac{(h_n\!+\!\omega_n)^2}{n^2}\right]^{2}
\qquad\Leftrightarrow \qquad e^{\omega_n t_n'}=\pm\left(\!\frac{h_n\!+\!\omega_n}{|n|}\!\right)^{2}
\qquad\qquad &\mbox{if } \omega_n\neq 0,\\[10pt] 
t_n'=\frac {2}{h_n}\qquad &\mbox{if } \omega_n= 0
\ea  %\label{maxpoint'} 
$$
(in fact, there is only one solution if $n$ is such that $\omega_n$ is real; 
the minus sign in the first line may occur only if  $\omega_n$ is imaginary).
Using (\ref{1dHn}) we find: \ 
$\dot H_n(t_n')=e^{-2} \!-\!2e^{-2} =-e^{-2}$  \ if $\omega_n= 0$, \  and 
\bea
&&\dot H_n(t_n')= \pm \frac {e^{-h_nt'_n}}{2\omega_n}\left[
(\omega_n\!+\!h_n)\frac{n^2}{(h_n\!+\!\omega_n)^2}\!+ \!(\omega_n\!-\!h_n)
\frac{(h_n\!+\!\omega_n)^2}{n^2}\right]\nn
&&= \pm\frac {e^{-h_nt'_n}}{2\omega_n}\frac {n^2\!\!-\!(h_n\!\!+\!\omega_n)^2}{h_n\!+\!\omega_n}
=\mp\frac {e^{-h_nt'_n}}{2\omega_n(h_n\!+\!\omega_n)}
\left[h_n^2\!+\!\omega_n^2\!+\!2h_n\omega_n \!-\! n^2\right]=\mp e^{-h_nt'_n}\nonumber
\eea
 if \  $\omega_n\neq 0$. It follows in either case (\ref{genineq'}).
Next, we show that
\beq
\ba{ll}
0<1\!-\!\dot H_n(t)<2h_n t\qquad\qquad &\mbox{if }\omega_n\ge 0, \\[8pt]    
0<1\!-\!\dot H_n(t) < 2h_n t\!+\!2 \sin^2\left(\!\frac{\Im({\omega_n})}2t\!\right)
\qquad\quad &\mbox{if }\omega_n=i\Im({\omega_n})\equiv i\sqrt{n^2\!-\!h_n^2}\in i\mathbb{R}.
\ea \label{buona}
\eeq
The first/third inequality in (\ref{buona}) was already proved in (\ref{genineq'}).
Using  (\ref{1dHn}) and $\cosh(\omega_nt)\ge 0$ we find
$1\!-\!\dot H_n(t)\le 1\!-\! e^{-h_nt}\!+\!h_nH_n(t)$; the second inequality then follows 
by (\ref{utile'}), (\ref{Hnineq}).
If $\omega_n\!\in\! i\mathbb{R}$ then (\ref{1dHn})  gives 
$$
1\!-\!\dot H_n(t)= 1\!-\! e^{-h_nt}\!+\!h_nH_n(t)\!+\!e^{-h_nt}2 \sin^2\!\left(\frac{\sqrt{n^2\!-\!h_n^2}}2t\right);
$$
this yields the fourth inequality  in (\ref{buona}) by (\ref{utile'}), (\ref{Hnineq}).
Finally, (\ref{buona'}) follows from (\ref{buona}) and \ $|\sin y|\le \min\{|y|,1\}$. \ \
\hfill \sq

\bigskip
{\it Proof of Proposition \ref{prop1}.} \ \ 
Using (\ref{disegHnl}) it is easy to check that for any 
$t>0$ the Fourier series (\ref{defvartheta}), all its term-by-term time derivatives
$\partial_t,\partial_t^2,...$, as well as  its term-by-term 
$\partial_x^2(\varepsilon\partial_t\!+\!\mbox{id}), 
\partial_x^2(\varepsilon\partial_t^2\!+\!\partial_t)$ derivatives,
converge absolutely and uniformly in $x$; consequently, (\ref{defvartheta}) defines
a continuous function $\vartheta(x,t)$ whose derivatives
$\vartheta_t,\vartheta_{tt},...,\partial_x^2(\varepsilon\vartheta_t\!+\!\vartheta)$,
$\partial_x^2(\varepsilon\vartheta_{tt}\!+\!\vartheta_t)$ are well-defined and
equal to the sum of these series, with $L\vartheta=L\vartheta_t=0$.
In fact, from (\ref{defvartheta})
$$
\ba{l}
2\pi\partial_t^l\vartheta(x,t)=\sum\limits_{n=-m}^{m}\frac {d^lH_n}{dt^l}e^{inx}+ R_m(x,t),
\qquad R_m^l(x,t):=\sum\limits_{ |n|>m}\frac {d^lH_n}{dt^l}e^{inx},\\[14pt]
|R_m^l(x,t)|\le \sum\limits_{ |n|>m}\left\vert\frac {d^lH_n}{dt^l}\right\vert\le 
\sum\limits_{ |n|>\bar n}\!
\frac {1}{2\omega_n} \left[\frac {2^l}{\varepsilon^l} \!+\! (a \!+\!\varepsilon n^2)^l
e^{-t \left(\varepsilon n^2-\frac 2\varepsilon\right) }\right]   
\ea   % \label{converge}
$$
($m\in\mathbb{N}$); the inequalites in the second line and the fact that
$\frac 1{\omega_n}\sim \frac 1{h_n}\sim \frac 1{n^2}$ \ as \ $|n|\to\pm\infty$ \
show that, for all  $t>0$ and $l=0,1,...$, the rest $R_m^l(x,t)$  goes to zero as $m\to\infty$ uniformly in $x$.
This shows the absolute and uniform (in $x$) convergence of the series  (\ref{defvartheta})
and all its term-by-term time derivatives. Similarly one proceeds for \
$\partial_x^2(\varepsilon\vartheta_t\!+\!\vartheta)$,
$\partial_x^2(\varepsilon\vartheta_{tt}\!+\!\vartheta_t)$.

Eq. (\ref{ineqn}) implies \ $\sum\limits_{n=1}^{\bar n}| H_n(t)| \le 1\!+\!
\sum\limits_{n=1}^{\bar n\!-\!1} \int\limits_n^{n\!+\!1}\!\! dy \frac 1 y = 
1\!+\!\int\limits_1^{\bar n}\!\! dy \frac 1 y=1\!+\!\log\bar n$ \ and
$$
2\pi|\vartheta(x,t)|\le 2\pi\vartheta(0,t)\le   \sum\limits_{n=1-\bar n}^{\bar n-1}
\left\vert H_n\right\vert+\!
\sum\limits_{ |n|\ge\bar n}\!\left\vert H_n\right\vert
\le \left\vert H_0\right\vert+2 +2\log\bar n
+\frac{2}{\varepsilon}\sum\limits_{ |n|\ge\bar n}\frac 1{n^2};   %    \label{converge'}
$$
the bound (\ref{thetaineq}) follows by  (\ref{defHn}),  (\ref{defbarn}) and
\ $\sum\limits_{ |n|\ge\bar n}\frac 1{n^2}\!\le\! 2\zeta(2)\!=\!\frac{\pi^2}{3}$,  \ 
a property of Riemann zeta function
\ $\zeta(s)\!:=\!\sum\limits_{n=1}^\infty\frac 1{n^s}$.  \
Definition (\ref{defbarn}) implies \
$\frac 2{\varepsilon}\!+\!1\ge\bar n\!>\! \frac 2{\varepsilon}$ \ and by (\ref{defHn})
\beq
\omega_n^2\ge n^2\!
\left(\!\frac{\varepsilon^2n^2}4\!-\!1\!\right)\ge \left\{\!
\ba{ll}
n^2\!\left(\!\frac{\varepsilon^2\bar n^2}4\!-\!1\!\right)> 3n^2, 
\qquad &\mbox{if }\: |n|\!\ge\! \bar n\\[10pt]
% n^2\!\left(\!\frac{\varepsilon^2\bar n^2|n|}4\!-\!1\!\right)> n^2(|n|\!-\!1)>(|n|\!-\!1)^3, 
%\qquad &\mbox{if }\: |n|\!\ge\! \bar n^2\\[10pt]
n^2\!\left(\!\frac{\varepsilon^2\bar n^2|n|^{3/2}}4\!-\!1\!\right)> n^2(|n|^{\frac 32}\!-\!1)>n^2|n\!\!-\!\!1|^{\frac 32},\qquad &\mbox{if }\:  |n|\!\ge\! \bar n^4.
\ea\right.  \label{omegabound}
\eeq
%The bound (\ref{thetatineq}) follows from  (\ref{defHn}),   (\ref{ineqn}) and
%$$
%\ba{l}
%2\pi |\vartheta_{t}(x,t) |\le\!
%\sum\limits_{n\in\mathbb{Z}}\!|\dot H_n(t)|\le 1\!\!+\!\!2\bar n^2\!+\!\!\!
%\sum\limits_{|n|>\bar n^2}\!\left[\!\frac 1{\varepsilon\omega_n}\!+
%\! 2e^{t \left(\!\frac 2\varepsilon\!-\!\varepsilon n^2\!
%\right) }\!\right] \le 1\!\!+\!\!2\bar n^2\!+\!\!\!
%\sum\limits_{|n|>\bar n^2}\!\frac 1{\varepsilon (n\!-\!1)^{3/2}}  \\[12pt]
%+ 2e^{\frac 2\varepsilon t}\! \sum\limits_{n\in\mathbb{Z}}\!\!e^{-\varepsilon n^2t }
%\le 1\!\!+\!\!2\bar n^2\!+\!\frac 1{\varepsilon}\zeta\!\!\left(\frac 32\right)\!+
%\! 2e^{\frac 2\varepsilon t}\theta\!\left(0,i\frac \pi\varepsilon t\right)
%\le 1\!\!+\!\!2\!\left(\!\frac 2{\varepsilon}\!+\!1\!\right)^2\! \!+\!\frac 3{\varepsilon}\!+
%\! 2e^{\frac 2\varepsilon t}\theta\!\left(0,i\frac \pi\varepsilon t\right), \ea
%$$
We prove (\ref{thetax}) using (\ref{disegHnl}-\ref{genineq'}), (\ref{genineq}), (\ref{omegabound}),
\ $\zeta(2)\!=\!\frac{\pi^2}6$, $\zeta(\frac 32)<3$ and $|n|\!\ge\! \bar n\:
\Rightarrow\:\frac{h_n}{\omega_n}\!\le\!2$: 
$$
\ba{l}
4\pi^2\Vert \vartheta_x(\cdot,\!t) \Vert_2^2=\sum\limits_{n\in\mathbb{Z}}\!| H\!_n\!(t)n|^2\le 2 \!\sum\limits_{n=1}^{\bar n}\! 1\!+\!2\! \sum\limits_{n=\bar n+1}^\infty
\!\frac {4}{\varepsilon^2 n^2}<2\bar n\!+\!\frac {8}{\varepsilon^2 } \sum\limits_{n=1}^\infty \frac 1{ n^2}<
2\!+\! \frac4{\varepsilon}\!+\! \frac{4\pi^2}{3\varepsilon^2}, \\[12pt]
4\pi^2\Vert \vartheta_{t}(\cdot,t) \Vert_2^2=\!
\sum\limits_{n\in\mathbb{Z}}\!\dot H_n^2(t)\le 1\!\!+\!\!2\bar n\!+\!\!\!
\sum\limits_{|n|>\bar n}\!\left[\!\frac 1{\varepsilon\omega_n}\!+
\! 2e^{t \left(\!\frac 2\varepsilon\!-\!\varepsilon n^2\!
\right) }\!\right]^2 \!\!\! \le 
1\!\!+\!\!2\bar n\!+\!\!\!
\sum\limits_{|n|>\bar n}\!\left[\!\frac 2{\varepsilon^2\omega_n^2}\!+
\! 8e^{2t \left(\!\frac 2\varepsilon\!-\!\varepsilon n^2\!
\right) }\!\right] \\[12pt]
\qquad\qquad\qquad \le\!3\!+\!\frac 4{\varepsilon}\!+\!\frac 4{3\varepsilon^2}
\!\!\sum\limits_{n=\bar n+1}^\infty\frac 1{n^2}
\!+\!  8  e^{\frac 4\varepsilon t}
\!\sum\limits_{n\in\mathbb{Z}} e^{-2\varepsilon n^2t }\!
<\! 3\!+\!\frac 4{\varepsilon}\!+\!\frac {2\pi^2}{9\varepsilon^2}\!+\!8e^{\frac {4t}\varepsilon}\,\theta\!\left(0,i\frac 2\pi\varepsilon t\right) \!, \\[12pt]
4 \pi^2\Vert \vartheta_{tx}(\cdot,t) \Vert_2^2=
\sum\limits_{n\in\mathbb{Z}}\!|\dot H\!_n\!(t)n|^2
\le \bar n^4(\bar n^4\!+\!1)\!+\!\!\! \sum\limits_{|n|>\bar n^4}\!n^2 \!\left[\!\frac 1{\omega_n\varepsilon}\!+
\!\frac {h_n}{\omega_n}e^{t \left(\!\frac 2\varepsilon-\varepsilon n^2\!
\right) }\!\right]^2\\
\qquad\qquad\qquad\:\:\le \bar n^4(\bar n^4\!+\!1)\!+\!\!\! \sum\limits_{|n|>\bar n^4}\!
\!\left[\!\frac {2n^2}{\omega_n^2\varepsilon^2}\!+
\!8n^2e^{2t \left(\!\frac 2\varepsilon-\varepsilon n^2\!
\right) }\!\right] \le \bar n^4(\bar n^4\!+\!1)\!+\!\frac 2{\varepsilon^2}\!\! 
\sum\limits_{|n|>\bar n^4}\! \!|n\!\!-\!\!1|^{-\frac 32} \\
\qquad\qquad\qquad\quad\:\:\!+\!16e^{\frac 4\varepsilon t}\sum\limits_{n\in\mathbb{Z}}\!\!n^2
e^{-2\varepsilon n^2t }\le  
\bar n^4(\bar n^4\!+\!1)\!+\!\frac 4{\varepsilon^2}\zeta\!\!\left(\frac 32\right)\!
-\!\frac 8\varepsilon \, e^{\frac 4\varepsilon t}\,
\partial_t\!\left[\theta\!\left(0,\!i\frac {2\varepsilon}\pi t\right)\right]. \hfill\sq 
\ea            
$$

\medskip
In spite of (\ref{thetax}), the Fourier series of
\ $\vartheta_x(\cdot, t)$ \  does not converge everywhere.
Moreover that of  $\vartheta_t$ diverges for $x=2k\pi$ and $t=0$.
However the Fourier series obtained deriving termwise $\vartheta$ (an arbitrary
number of times) w.r.t. $x,t$ define for all $(x,t)\in\mathbb{R}\times I$ 
(time-dependent) periodic distributions (see e.g. \cite{Bea73}) w.r.t. the
$x$ variable, since 
their coefficients grow {\it slowly} with $|n|$ (i.e. at most with a 
power law), by  (\ref{disegHnl}).
The space $S$ of test functions consists of infinitely differentiable periodic functions.
The coefficients $g_n$ of the Fourier expansion
\beq
g(x)=\sum\limits_{n\in\mathbb{Z}} g_n e^{inx},
\qquad\qquad g_n :=\frac 1{2\pi}\int\limits_0^{2\pi} \!\! dx g(x)e^{-inx}, \label{Fourier}
\eeq
 of \ $g\!\in\! S \ $have a {\it fast} decrease with $|n|$  (i.e. faster than any power). 
By definition, applying $\eta\in S'$ (the space of periodic distributions) to a $g\in S$ 
gives
\beq
\langle \eta, g\rangle=\sum\limits_{n\in\mathbb{Z}} \eta_{-n} g_n.   \label{functional}
\eeq
As said,  \ $H_n\!\!\to\! 0$,    $\dot H_n\!\!\to\! 1$ \ as $t\!\to\! 0$.\  Hence
\beq
\lim\limits_{t\downarrow 0}\vartheta(x,t)=0,\qquad\qquad
\lim\limits_{t\downarrow 0}\vartheta_t(x,t)= \sum\limits_{k\in\mathbb{Z}}\delta(x\!-\!2\pi k)
                                                     \label{deltalimit}
\eeq
 in the sense of convergence in $S'$; the rhs(\ref{deltalimit})$_2$
is the periodic delta function.
In fact, $\vartheta$  is the only $t$-dependent periodic
distribution fulfilling (\ref{thetaprop}), (\ref{deltalimit}).

We can and shall use (\ref{functional}) as a definition of functional $\eta$ also on
less regular functions spaces. For instance, if  $g\!\!\in\!\! L^2([0,\!2\pi])$ then  
(\ref{functional}) makes sense for $\eta\!\!\in\!\! L^2([0,\!2\pi])$, and
\beq
\langle \eta, g\rangle=\int_0^{2\pi} \!\! \frac {dx}{2\pi}\, \eta(x) g(x).
\eeq

The fundamental solution $K(x,t)$ of the equation $Lu=0$ on $\mathbb{R}\!\times\!\mathbb{R}^+$ 
was determined for $ a \!=\!0$ in  \cite{DacRen92} and for $ a \!>\!0$ in 
\cite{DacDeaRen97} in the form of quite complicated integrals involving the modified
Bessel function of order  zero. \ $K(\cdot,t)$
is a Schwarz function for any $t\!>\!0$. Since \ $K(x,t)\!\to\! 0$, \  
$K_t(x,t)\!\to\!\delta(x)$ \ (in the sense of convergence of tempered distributions) 
as  $t\to 0$, the present  $\vartheta$ must be related to $K$ by
\beq
\vartheta(x,t) =   \sum_{m\in\mathbb{Z}}K(x\!+\!2m,t).            \label{thetaK}
\eeq
This is the analog of a property of Jacobi Theta function,
which is the fundamental solution of the heat equation. Eq. (\ref{thetaK}) was used
as a {\it definition} of $\vartheta$ in the DBC case for $ a \!=\!0$  in  
\cite{DacDan98,DanFio05}. However, here we prefer to work directly with the simpler 
and explicit definition (\ref{defvartheta}), as done in \cite{Dea01} for the
DBC case (alone).

\section{Green functions and convolutions with them}
\label{Green}

We define the Green functions appropriate for the three boundary conditions:
\beq                                                   \label{Greenf}
\ba{ll}
w^p(x,t;\xi):=
\vartheta(x\!-\!\xi,t)=\frac 1{2\pi}\sum\limits_{n\in\mathbb{Z}} H_n(t)e^{in(x-\xi)}
\qquad\qquad &\mbox{PBC},\\[6pt]
w^d(x,t;\xi):=\vartheta(x\!-\!\xi,t)-\vartheta(x\!+\!\xi,t)\!=\!
\frac 2{\pi}\sum\limits_{n=1}^\infty\! H_n(t)
\sin(nx)\sin(n\xi)
\qquad &\mbox{DBC},\\[6pt]
w^n\!(x\!,\!t\!;\!\xi):=\vartheta(x\!-\!\xi,\!t)\!+\!\vartheta(x\!+\!\xi,\!t)\!=
\!\frac {1\!-\!e^{- a  t}}{\pi a }\!+\!
\frac 2{\pi}\!\!\sum\limits_{n=1}^\infty\!\!\! H_n\!(t)
\cos(nx)\cos(n\xi)\qquad &\mbox{NBC}.
\ea
 \eeq
For $t\!>\!0$ \ and \ $w=w^p,w^d,w^n$ \ 
$w$ and the derivatives $w_t,w_{tt},w_{ttt}$, $\partial_x^2(\varepsilon w_t\!+\!w)$, 
$\partial_x^2(\varepsilon w_{tt}\!+\!w_t)$, $\partial_\xi^2(\varepsilon w_t\!+\!w)$
%$\partial_\xi^2(\varepsilon w_{tt}\!+\!w_t)$
are continuous functions of $x,t,\xi$,  of period $2\pi$ w.r.t. both variables $x,\xi$, fulfilling
\beq
Lw=0, \qquad\qquad Lw_t=0.
\label{wxx=wxixi}
\eeq
For all $t$ they are $t$-dependent distributions w.r.t. both variables $x,\xi$ fulfilling
 \beq
\ba{lr}                      \label{wbc}
%w^p(x,k\pi,t)=\vartheta(x\!-\!k\pi,t),\qquad & \mbox{PBC},\\[8pt]
w^d(k\pi,t;\xi)=w^d(x,t;k\pi)\equiv 0,\qquad\qquad & %w^d_x(x,k\pi,t)=-2\vartheta_x(x\!-\!k\pi,t),\\[8pt]
%w^n(x,k\pi,t)=2\vartheta(x\!-\!k\pi,t),\qquad &
w^n_x(k\pi,t;\xi)=w^n_x(x,t;k\pi)\equiv 0
\ea 
%w^p_x(x,k\pi,t)=-\vartheta_x(x\!-\!k\pi,t)\qquad\qquad & \mbox{PBC},\\[6pt]
\eeq
for all \ $k\in\!\mathbb{Z}$, \
and analogous relations obtained deriving both sides w.r.t. $t$.
Moreover, \ $w=w^p,w^d,w^n$ \ fulfill limits analogous to (\ref{deltalimit}).

Let $C^{pk}$ be the space of (complex) functions of period $2\pi$ continuous with
their derivatives up to the $k$-th order. If  $g\in C^{p0}$ then
$g\in L^2([0,2\pi])$ as well, what implies  \
$\Vert g\Vert_2^2=\sum\limits_{n\in\mathbb{Z}} |g_n|^2<\infty$; \
whereas if $g\in C^{p1}$ then, as known, \
$\Vert g\Vert_1=\sum\limits_{n\in\mathbb{Z}} |g_n|<\infty$, \ and the series (\ref{Fourier})
converges absolutely and uniformly to $g(x)$ in all of $\mathbb{R}$. \
Similarly for \  $g\in C^{pk}$  \ with  \ $k>1$. \ Let  
\bea
&& C^{dk}:=\{g\in  C^{pk}\:|\: g(x)=g(\pi)=0\},\qquad 
C^{nk}:=\{g\in  C^{pk}\:|\: g_x(x)=g_x(\pi)=0\}, \nn[10pt]
&& \ba{lll}
w(x,t;\xi)=w^p(x,t;\xi), \qquad\qquad & {{\sf D}}=[0,2\pi],
\qquad\qquad &\mbox{\rm PBC},\\[6pt]
w(x,t;\xi)=w^d(x,t;\xi),\qquad\qquad & {{\sf D}}=[0,\pi],
\qquad &\mbox{\rm DBC},\\[6pt]
w(x,t;\xi)=w^n(x,t;\xi), \qquad\qquad & {{\sf D}}=[0,\pi],
\qquad\qquad &\mbox{\rm NBC}.
\ea   \nonumber
\eea
For \ $w=w^p,w^d,w^n$ \ and resp. \ $g\in  C^{p0}, C^{d0}, C^{n1}$ \ let
\beq
 w^g(x,t):=\langle w(x,t;\cdot),g\rangle=\int_{\sf D} \!\! d\xi\:w(x,t;\xi) g(\xi).
%,\qquad\quad w_t^g(x,t):=\int\limits_{\sf D} \!\! d\xi\:w_t(x,t;\xi) g(\xi).  
\label{wg}
\eeq
 $w^{pg}$ is just the convolution of $\vartheta, g$. By a straightforward calculation
\beq
\ba{ll}
 w^{pg}(x,t)=%\int_0^{2\pi} \!\! d\xi\:\vartheta(x\!-\!t;\xi) g(\xi)=
\sum\limits_{n\in\mathbb{Z}} H_n(t)e^{inx}g_n,\qquad\qquad & g\in  C^{p0},\\[10pt]
w^{dg}(x,t)=\sum\limits_{n=1}^\infty H_n(t) \sin(nx)g_n,\qquad\qquad & g\in  C^{d0},\\[10pt]
w^{ng}(x,t)=\frac {1\!-\!e^{- a  t}}{ a }g_0+
\sum\limits_{n=1}^\infty H_n(t) \cos(nx)g_n, \qquad\qquad & g\in  C^{n1}.
\ea                           \label{wg'}
\eeq

\begin{prop}
For  \ $w^g=w^{pg},w^{dg},w^{ng}$ \  
(with $g\in C^{p0},C^{d0}, C^{n1}$ respectively)
the series (\ref{wg'}) define for all \ $t\ge 0$ \ 
a continuous function; $w^g$ is real-valued if $g$ is.
For \ $t\!>\!0$ \ the derivatives  $w^g,w^g_t,w^g_{tt},w^g_{ttt}$, 
$\partial_x^2(\varepsilon w^g_t\!+\!w^g)$, $\partial_x^2(\varepsilon w^g_{tt}\!+\!w^g_t)$
are well-defined, are uniformly in $x$ the sum of the 
corresponding term-by-term derived series and fulfill
\beq
 Lw^g=0,\qquad\qquad  Lw^g_t=0.                       \label{Lwg}
\eeq
$w^g,w^g_t$ \ fulfill the `initial' conditions
\bea
 &&\lim\limits_{t\downarrow 0}w^g{}(x,t)\equiv 0
\qquad\qquad\quad\:\mbox{uniformly in }x,\label{wg0}\\
&&\lim\limits_{t\downarrow 0}w^g_t(x,t)=g(x)\qquad\qquad \mbox{uniformly in }x\quad 
\mbox{if }g\in C^1   \label{wgt0}
\eea
and the respective boundary conditions, more precisely:
\beq
\ba{ll}
w^{pg}(\cdot,\!t)\in  C^{p1},\qquad w^{dg}(\cdot,\!t)\in  C^{d1},\qquad
w^{ng}(\cdot,\!t)\in  C^{n2},\qquad\quad &   t\ge 0,          \\[8pt]
w^{pg}_t(\cdot,\!t)\in  C^{p1},\qquad w^{dg}_t(\cdot,\!t)\in  C^{d1},\qquad
w^{ng}_t(\cdot,\!t)\in  C^{n2},\qquad\quad & t> 0.
\ea\label{wg''}
\eeq
If in addition \ $g$ \ has a continuous second derivative, then 
\beq
 %\int_{\sf D}\!\!\!\!d\xi\uo(\xi)\, w_{tt}(x,t;\xi)=
\int_{\sf D}\!\!\!\!d\xi\, g(\xi)\,[\partial_x^2(\varepsilon w_t\!+\!w)](x,t;\xi)
=\int_{\sf D}\!\!\!\!d\xi\, g''(\xi)\,[\varepsilon w_t\!+\!w](x,t;\xi). \label{byparts}
\eeq
\label{prop2}
\end{prop}

The results in Proposition \ref{prop2} generalize results of \cite{Dea01}.

The regularity of $w^g$ improves with that of $g$;
in particular, if $g$ is infinitely differentiable, so is $w^g$.
If $g\notin C^1$, 
for (\ref{wgt0}) to hold at $x$ it suffices that left and right derivatives of 
$g$ both exist at $x$, by standard wisdom about the Fourier series.

\medskip
{\it Proof.} \ \ The mentioned series converge uniformly in $x$ for
 $t\!>\!0$ by
(\ref{disegHnl}),  (\ref{thetax}) and  Schwarz inequality in $l^2(\mathbb{Z})$. \
Eq. (\ref{Lwg}) follows from (\ref{thetaprop}).
One can check (\ref{byparts}) just by noting that the Fourier expansions
of both sides coincide and converge, see (\ref{wg'}).
To prove  (\ref{wg''}) first note that each term of the Fourier
series (\ref{wg'}) fulfills the corresponding boundary conditions,
then that the series and their term-by-term derivatives converge uniformly.
As examples we prove (\ref{wg''})$_1$, (\ref{wg''})$_4$ : \
$w^{pg}_x(x,\!t)\!=\!i\!\sum\limits_{n\in\mathbb{Z}}\!H\!_n\!(t)g_n n e^{inx}$, \
$w^{pg}_{tx}(x,\!t)\!=\!i\!\sum\limits_{n\in\mathbb{Z}}\!\dot H\!_n\!(t)g_n n e^{inx}$; \
using   Schwarz inequality in $l^2(\mathbb{Z})$ and (\ref{thetax}) we find
\bea
\left|w^{pg}_x\!(x,\!t)\right| &\le& \sum\limits_{n\in\mathbb{Z}}\!| H\!_n\!(t)g_n n|\le 
\!\left[\sum\limits_{n\in\mathbb{Z}}\!|H\!_n\!(t)n|^2\!\right]^{\frac 12}
\!\!\left[\sum\limits_{n\in\mathbb{Z}}\!|g_n|^2\!\right]^{\frac 12}\nn[6pt]
&=& 2\pi\Vert \vartheta_x(\cdot,t) \Vert_2 \,\Vert g \Vert_2<
\sqrt{2\!+\! \frac4{\varepsilon}\!+\! \frac{2\pi^2}{3\varepsilon}} \,\Vert g \Vert_2 <\infty 
\qquad\qquad t\ge 0,\nn[8pt]
\left|w^{pg}_{tx}\!(x,\!t)\right| &\le& \sum\limits_{n\in\mathbb{Z}}\!| \dot H\!_n\!(t)g_n n|\le 
\!\left[\sum\limits_{n\in\mathbb{Z}}\!|\dot H\!_n\!(t)n|^2\!\right]^{\frac 12}
\!\!\left[\sum\limits_{n\in\mathbb{Z}}\!|g_n|^2\!\right]^{\frac 12}\nn[6pt]
&=& 2\pi\Vert \vartheta_{tx}(\cdot,t) \Vert_2 \,\Vert g \Vert_2<\infty
\qquad\qquad t> 0.
\nonumber
\eea

Next, from (\ref{Hnineq}) and  (\ref{thetaineq}) it follows
\beq
4\pi^2\Vert \vartheta(\cdot, t) \Vert^2_2\equiv\sum\limits_{n\in\mathbb{Z}}|H_n(t)|^2
\le t\sum\limits_{n\in\mathbb{Z}}|H_n(t)|\le t \, N(t)    \label{theta2ineq}
\eeq
 for all $t$. Therefore not only the sequence $\{H_n(t)\}_{n\in\mathbb{Z}}$ is in $l^2(\mathbb{Z})$
for all $t$, but its norm goes to 0 as $t\to 0$. Using again  Schwarz inequality in 
$l^2(\mathbb{Z})$ we find as a consequence
$$
\left|w^{pg}(x,t)\right|\le \sum\limits_{n\in\mathbb{Z}}| H_n\!(t)g_n|\le 
\left[\sum\limits_{n\in\mathbb{Z}}\!|H_n\!(t)|^2\right]^{\frac 12}\!\!\left[\sum\limits_{n\in\mathbb{Z}}\!|g_n|^2\right]^{\frac 12}\!\!\!< \sqrt{t\, N(t)}\,\Vert g\Vert_2 . 
$$
This shows (\ref{wg0}) in the PBC case.  On the other hand, for any \  $m\!\in\!\mathbb{N}$  \  it is 
\bea
&&\left|g(x,t)\!-\!w^{pg}_t\!(x,t)\right| \le \sum\limits_{n=-m}^m\!\left|1\!-\!\dot H_n(t)\right||g_n|+
\sum\limits_{|n|\!>\!m}\!\left|1\!-\!\dot H_n(t)\right||g_n|\nn
&&\le t\left[\!\sum\limits_{n=-m}^m\!\!2h_n
|g_n|\!+\!\sum\limits_{n\in\mathbb{Z}: \atop\omega_n\in i\mathbb{R}}|\Im({\omega_n})||g_n|\right]\!
+2\sum\limits_{|n|\!>\!m}\!\!|g_n|<
t\left(2h_m\!+\!n_+\right)\Vert g\Vert_1+2\sum\limits_{|n|\!>\!m}\!\!|g_n|,\nonumber
\eea
where we have used the inequality $|\Im({\omega_n})|\!<\!|n|\!<\!n_+$ following from (\ref{meq}) 
and the fact that the sequence $\{h_n\}_{n\in\mathbb{N}}$ is increasing.
For any \ $\eta\!>\!0$ \ choose \ $m$ \ so large that 
\ $2\!\sum_{|n|\!>\!m}\!\!|g_n|<\eta/2$. 
\ Setting \  $\delta\!:=\!\eta/2(2h_m\!+\!n_+)\Vert g\Vert_1$
we find
\beq
\left|g(x,t)\!-\!w^{pg}_t\!(x,t)\right|<\eta\qquad \qquad\forall\, x\in\mathbb{R},\quad t<\delta.
\eeq
This shows (\ref{wgt0}) in the PBC case. 

Applying to $g^d\in  C^{d0}$ and $g^n\in  C^{n1}$ respectively  odd, even extensions 
defined as in (\ref{evenoddext}-\ref{evenoddext'}) (ignoring the $t$ dependence there) one 
obtains 
%setting
%$$
%\ba {ll}
%\tilde g_n:=\left\{\ba {ll}\mbox{sign}(n)\, \frac 1{2i} g_n\qquad\qquad &\mbox{if } n\neq 0,\\[6pt]
%0 \qquad\qquad &\mbox{if } n=0,\ea\right.\qquad\qquad & g\in  C^{d0},\\[12pt]
%\tilde g_n:=\left\{\ba {ll}\frac 1{2} g_n\qquad\qquad\qquad\qquad &\mbox{if } n\neq 0,\\[6pt]
%g_0 \qquad\qquad\qquad\qquad &\mbox{if } n=0\ea,\right.\qquad\qquad & g\in  C^{n0},
%\ea
%$$ defines 
a $g^p\in  C^{p0}$ such that \  $w^{dg^d}=w^{pg^p}$ in $[0,\pi]$ \ 
and a $g^p\in  C^{p1}$  such that \  $w^{ng^n}=w^{pg^p}$ in $[0,\pi]$, \ respectively.
This shows that  (\ref{wg''}),  (\ref{wg0}-\ref{wgt0}) in the DBC, NBC cases follow
from  (\ref{wg''}),  (\ref{wg0}-\ref{wgt0}) in the PBC case.  \hfill \sq

\section{Existence and uniqueness}
\label{exist}

\begin{prop} Problem  (\ref{22'}-\ref{23'}) is  equivalent to the integral equation
\bea
u(x,t)&=&\int_{\sf D}\!\!\!\!d\xi \left\{
\Big[\uu\!-\!\varepsilon \uo''\!+\! a  \uo \Big]\!(\xi)\,
w(x,t;\xi)\!+\!\uo(\xi)\, w_t(x,t;\xi)\right\}\nn
&&  +\int_0^{t}\!\!\!\!d\tau\!\int_{\sf D}\!\!\!\! d\xi\:
f\!\Big[\!\xi,\tau,U(\xi\!,\!\tau)\!\Big]w(x,t\!-\!\tau;\xi).            \label{inteq'}
%+2\pi m\!\left[\varepsilon \vartheta_x(x,t)
%\!+\!\int_0^t\!\!\!dt'\,\vartheta_x(x,t')\right].
\eea
\end{prop}

{\it Proof.} \ \
Let \ $L'\!:=\!\partial_\tau^2\!-\! a \partial_\tau\!-\!\partial_\xi^2(1\!-\!\varepsilon
\partial_\tau)$. \ Assuming that $u(x,t)$ solves  (\ref{22'})$_1$
it is straightforward to prove the identity \cite{DacDan98}
 \beq                                                   \label{24}
\partial_\xi (u	\tilde w_\xi\!-\!u_\xi \tilde w\!+\!\varepsilon u_\xi \tilde w_\tau\!-\!
\varepsilon u\tilde w_{\xi\tau})\!+\!\partial_\tau(u_\tau \tilde w\!-\!\varepsilon u_{\xi\xi}\tilde w
\!+\! a  u \tilde w\!-\!u\tilde w_\tau)-f\tilde w+u L'\tilde w=0,\qquad
 \eeq
for  any smooth functions \  $u(\xi,\tau)$,  $\tilde w(\xi,\tau)$. \ 
Choosing \ $\tilde w(\xi,\tau)=w(x,t\!-\tau;\xi)$, \  with \  $w=w^p,w^d,w^n$ \ resp. in
the PBC, DBC, NBC cases, the term $u L'\tilde w$ becomes identically zero. Integrating
(\ref{24}) in $d\xi$ over the respective domains ${\sf D}$ we obtain
\bea
&&0=\int_{\sf D}\!\!d\xi\left[\partial_\xi(u \tilde w_\xi\!-\!u_\xi  \tilde w\!+\!
\varepsilon u_\xi  \tilde w_\tau\!-\!\varepsilon u \tilde w_{\xi\tau})
\!+\!\partial_\tau(u_\tau  \tilde w\!-\!\varepsilon u_{\xi\xi}\tilde w \!+\! 
a  u  \tilde w\!-\!u \tilde w_\tau)-f\tilde  w\right]\nn
&&=\left[u \tilde w_\xi\!-\!u_\xi  \tilde w\!+\!\varepsilon u_\xi  \tilde w_\tau\!-\!\varepsilon
u \tilde w_{\xi\tau}\right]^{\xi=b}_{\xi=0}\!+\!\int_{\sf D}\!\!d\xi\left[\partial_\tau
(u_\tau  \tilde w\!-\!\varepsilon u_{\xi\xi} \tilde w\!+\!
 a  u  \tilde w\!-\!u \tilde w_\tau)-f \tilde w\right],     \nonumber
\eea
where $b=2\pi$ in the PBC case and  $b=\pi$ in the DBC, NBC cases.
The expression in the square bracket vanishes in all three cases
by the periodicity of $w^p,w^p_\tau,w^p_\xi, w^p_{\xi \tau}$ w.r.t. to
$\xi$ or the boundary conditions (\ref{23'}), (\ref{wbc}).
By further integrating in $d\tau$ over $]\eta,t\!-\!\eta[$ (where
$\eta>0$) we find
\bea
&&\int_\eta^{t\!-\!\eta}\!\!\!\!\!\!d\tau\!
\int_{\sf D}\!\!\!\!d\xi\: f\tilde w= \int_{\sf D}\!\!\!\!d\xi\, \left[u_\tau \tilde w\!-\!\varepsilon u_{\xi\xi}\tilde w\!+\! a  u \tilde w\!-\!u\tilde w_\tau \right]_{\tau\!=\!\eta}^{\tau\!=\!t\!-\!\eta} \nn &&
=\int_{\sf D}\!\!\!\!d\xi \Big\{[u_\tau \!\!-\!\!\varepsilon u_{\xi\xi}\!+\! a  u]\!(\xi,\!t\!\!-\!\!\eta)  
w(x,\!\eta;\!\xi)\!-\!u(\xi,\!t\!\!-\!\!\eta) w_\tau \!(x,\!\eta;\!\xi)\nn 
&& \qquad \quad - [u_\tau \!\!-\!\!\varepsilon u_{\xi\xi}\!+\! a  u]\!(\xi,\!\eta)  
w(x,\!t\!\!-\!\!\eta;\!\xi)\!-\!u(\xi,\!\eta) w_\tau \!(x,\!t\!\!-\!\!\eta;\!\xi)
\Big\}.                     \nonumber
\eea
By Schwarz inequality
$$
\left|\int_{\sf D}\!\!\!\!d\xi\, [u_\tau \!\!-\!\!\varepsilon u_{\xi\xi}\!+\! a  u]\!(\xi,\!t\!\!-\!\!\eta)w(x,\!\xi,\!\eta)  \right|
\le \Vert [u_\tau \!\!-\!\!\varepsilon u_{\xi\xi}\!+\! a  u]\!(\cdot,\!t\!\!-\!\!\eta)\Vert_2\:
\Vert \vartheta(\cdot,\!\eta)\Vert_2;
$$
by (\ref{theta2ineq}), this goes to zero as $\eta\to 0$. 
\ Letting $\eta\rightarrow 0$, by  (\ref{22'}), (\ref{wgt0}) we find that $u$ satisfies the integral 
equation (\ref{inteq'}).

\medskip
Conversely, assume $u$ solves (\ref{inteq'}). 
The rhs of the first line of (\ref{inteq'}) is nothing but \ 
$w^{\uu\!+\! a  \uo \!-\!\varepsilon \uo''}+w_t^{\uo}$; \ 
it fulfills the respective boundary conditions by (\ref{wg''}).
Also the second line fulfills the respective boundary conditions, by 
(\ref{wbc}). Next, let us check  that  $u$  fulfills $Lu=f$.
$L$ applied to the first line of (\ref{inteq'}) gives zero, by (\ref{Lwg}).
Denoting as ${\cal I}$ the second line, we find
\bea
{\cal I}_t&=&\int_{\sf D}\!\!\!\! d\xi\: f\!\Big[\!\xi,t,U
(\xi\!,\!t)\!\Big]w(x,0;\xi)   +\int_0^{t}\!\!\!\!d\tau\!\int_{\sf D}\!\!\!\! d\xi\:
f\Big[\xi,\tau,U(\xi\!,\!\tau)\Big]w_t(x,t\!-\!\tau;\xi)\nn
&=&\int_0^{t}\!\!\!\!d\tau\!\int_{\sf D}\!\!\!\! d\xi\:
f\Big[\xi,\tau,U(\xi\!,\!\tau)\Big]w_t(x,t\!-\!\tau;\xi)  \label{It}\\
{\cal I}_{tt}&=&\int_{\sf D}\!\!\!\! d\xi\: f\Big[\xi,t,U(\xi\!,\!t)\Big]w_t(x,0;\xi)   +
\int_0^{t}\!\!\!\!d\tau\!\int_{\sf D}\!\!\!\! d\xi\:
f\Big[\xi,\tau,U(\xi\!,\!\tau)\Big]w_{tt}(x,t\!-\!\tau;\xi)\nn
&=& f\Big[x,\!t,\!U(x,\!t)\Big]+
\int_0^{t}\!\!\!\!d\tau\!\int_{\sf D}\!\!\!\! d\xi\:
f\Big[\xi,\tau,U(\xi\!,\!\tau)\Big]w_{tt}(x,t\!-\!\tau;\xi)\qquad\quad\Rightarrow\nn
Lu&=& L{\cal I}= f\Big[x,\!t,\!U(x,\!t)\Big]+
\int_0^{t}\!\!\!\!d\tau\!\int_{\sf D}\!\!\!\! d\xi\:
f\Big[\xi,\tau,U(\xi\!,\!\tau)\Big](Lw)(x,t\!-\!\tau;\xi)\nn
&=& f\Big[x,\!t,\!U(x,\!t)\Big]
\eea
as claimed. We have used  (\ref{wg0})   in the second equality,
  (\ref{wgt0})  in the fourth,  (\ref{Lwg}) in the sixth.
Finally, let us check that $u$ fulfills the required initial conditions. 
Taking the limit $t\downarrow 0$ and using (\ref{wg0}), (\ref{wgt0})
it is straightforward to show that (\ref{inteq'}) implies \ $u(x,0)=\uo(x)$. \ 
We now evaluate $u_t(x,t)$:
\bea
&& u_t(x,t)=\int_{\sf D}\!\!\!\!d\xi \left\{
\Big[\uu\!+\! a  \uo \!-\!\varepsilon \uo''\Big]\!(\xi)\,
w_t(x,t;\xi)\!+\!\uo(\xi)\, w_{tt}(x,t;\xi)\right\}+{\cal I}_t  \nn
&& =\int_{\sf D}\!\!\!\!d\xi \left\{\!
\Big[\uu\!+\! a  \uo \!-\!\varepsilon \uo''\Big]\!(\xi)\,
w_t(x,t;\xi)\!+\!\uo(\xi)\, [\partial_x^2(\varepsilon w_t\!+\!w)\!-\!aw_t]
(x,t;\xi)\!\right\}+{\cal I}_t\nn
&& =\int_{\sf D}\!\!\!\!d\xi \left\{
\Big[\uu \!-\!\varepsilon \uo''\Big]\!(\xi)\,
w_t(x,t;\xi)+ \uo''(\xi)\,[\varepsilon w_t\!+\!w](x,t;\xi)\right\} + {\cal I}_t \nn
&& =\!\int_{\sf D}\!\!\!\!d\xi \left[ \uu\!(\xi) 
w_t(x,\!t;\!\xi)\!+\!\uo''(\xi)w(x,\!t;\!\xi)
\right] +\!\int_0^{t}\!\!\!\!\!d\tau\!\!\int_{\sf D}\!\!\!\! d\xi\,
f\!\Big[\!\xi,\!\tau,\!U(\xi\!,\!\tau)\!\Big]w_t(x,\!t\!-\!\tau;\!\xi);
\nonumber  
\eea
we have used  (\ref{It})  in the  first equality,  (\ref{wxx=wxixi})  in the  second, 
 (\ref{byparts}) in the third.
Taking the limit $t\downarrow 0$ and using (\ref{wg0}), (\ref{wgt0})
we find  $u_t(x,0)=\uu(x)$, as claimed. \hfill \sq

\medskip
If $f=f(x,t)$, the rhs(\ref{inteq'}) gives the unique explicit
 solution of (\ref{22'}-\ref{23'}).
Otherwise, to deal with the integro-differential equation (\ref{inteq'})
it is convenient to reformulate it in any finite time interval
$[0,T]$ as the fixed point equation
\beq
\T u=u.
\eeq
$\T$ is the linear map $\T: \B\mapsto \B$ defined by
\beq
\ba{l}
\B:=\!\{v(x,t)\:\mbox{of period }2\pi\:\:|\:\: v,v_x,v_t\!\in\! C(D_T)\},
\qquad D_T\!:=\!{\sf D}\!\times\![0,T]\\[8pt]
[\T v](x,t):=\displaystyle\int_{\sf D}\!\!\!\!d\xi \left\{
\Big[\uu\!+\! a  \uo \!-\!\varepsilon \uo''\Big]\!(\xi)\,
w(x,t;\xi)\!+\!\uo(\xi)w_t(x,t;\xi)\right\}\\[8pt]
\qquad\qquad\qquad +\displaystyle\int_0^{t}\!\!\!\!d\tau\!
\displaystyle\int_{\sf D}\!\!\! d\xi\: f\!\Big[\!\xi,\tau,V
(\xi\!,\!\tau)\!\Big]w(x,t\!-\!\tau;\xi).
\ea
\eeq
$\B$ is a Banach space  w.r.t. the norm 
\beq
 \|v\|_{\lambda, {\scriptscriptstyle T}}:=\max_{D_T}|{\rm e}^{-\lambda t}v(x,\!t)|\!
 +\!\max_{D_T}|{\rm e}^{-\lambda t}v_x(x,\!t)|
\!+\! \max_{D_T}|{\rm e}^{-\lambda t}v_t(x,\!t)|,
\eeq
where $\lambda$ is some positive constant we fix below. 
We shall assume (in all three cases) that $f$ is continuous in 
$(x,t,v)\in D\times I \times\mathbb{R}^3$  and satisfies a Lipschitz condition w.r.t. 
$v_1,v_2,v_3%,v_4
$:
\beq
|f(x,t,v)-f(x,t,y)| \leq \mu \big(|v_1\!-\!y_1|\!+\!|v_2\!-\!y_2|\!+\!|v_3\!-\!y_3|\big), 
\qquad\mu\in\mathbb{R}^+.                         \label{Lipsch}
\eeq
Note that (\ref{Lipsch}) remains true for $\hat f$ after the transformations $f\mapsto \hat f$
defined in (\ref{redef'}) and  $f\mapsto \tilde f$ defined in (\ref{redef0}).
We can now state the main result of the present paper.

\begin{theorem} 
If \ $f=f(x,t,v)$ \ 
is continuous and Lipschitz with respect to $v_1,v_2,v_3$, then the
 nonlinear problem (\ref{eqh}) with Dirichlet, Neumann, or pseudoperiodic
 boundary conditions (\ref{23})   has a unique  solution in all $D\times[0,\infty[$.
\end{theorem} 

{\it Proof.} \ \ 
If $a\!<\!0$ we apply the change of variables (\ref{redef0})  and reduce 
the existence and uniqueness theorem for the case $a<0$ to the one for  the case 
 $a=0$. So it suffices to prove the theorem for $a\ge 0$.
Let\  $\Delta\! f\!(\xi\!,\!\tau)\!:=\!f\!\big[\!\xi\!,\!\tau\!,V_1(\xi\!,\!\tau)\!\big]\!-\!
f\!\big[\!\xi\!,\!\tau\!,V_2(\xi\!,\!\tau)\big]$. \ Using (\ref{Lipsch}) we find for \ $(\xi,\tau)\!\in\!D_T$
\bea
%\Delta\! f\!(\xi\!,\!\tau)\!:=\!f\!\big[\!\xi\!,\!\tau\!,V_1(\xi\!,\!\tau)\!\big]\!-\!
%f\!\big[\!\xi\!,\!\tau\!,V_2(\xi\!,\!\tau)\big]
%= \sum\limits_{n\in\mathbb{Z}}  \Delta\! f_n(\tau)\,e^{in\xi}
% \label{defDeltaf}
%\nn \stackrel{(\ref{Lipsch})}{\Longrightarrow} \qquad\qquad
|\Delta\! f\!(\xi\!,\!\tau)|e^{-\lambda \tau}\le \mu \Vert v_1\!-\!v_2\Vert_{\lambda, {\scriptscriptstyle T}}, \qquad\quad
\Vert \Delta\! f(\cdot,\!\tau)\Vert_2^2=
%\sum\limits_{n\in\mathbb{Z}}  \Delta\! f_n^2(\tau)=
\displaystyle\int_{\sf D}\!\frac{d\xi}{ 2\pi}\:\Delta\! f^2\!(\xi\!,\!\tau)\le
\mu^2\Vert v_1\!-\!v_2\Vert_{\lambda, {\scriptscriptstyle T}}^2\, e^{2\lambda \tau}\!. \label{Deltaf}
\eea
From (\ref{inteq'}) and (\ref{wg0}) we obtain for \ $(x,t)\!\in\!D_T$
\beq
\ba{lll}
[{\cal T}\!v_1\!-\!{\cal T}\!v_2](x,t) &=& \displaystyle\int_0^{t}\!\!\!\!d\tau \!\!\!
\displaystyle\int_{\sf D}\!\!\!\!d\xi\: w(x,t\!\!-\!\!\tau;\xi)\, \Delta\! f\!(\xi\!,\!\tau),\\[8pt]
[{\cal T}\!v_1\!-\!{\cal T}\!v_2]_x(x,t) &=& \displaystyle\int_0^{t}\!\!\!\!d\tau \!\!\!
\displaystyle\int_{\sf D}\!\!\!\!d\xi\: w_x(x,t\!\!-\!\!\tau;\xi)\, \Delta\! f\!(\xi\!,\!\tau),\\[8pt]
[{\cal T}\!v_1\!-\!{\cal T}\!v_2]_t(x,t)  & = & \displaystyle\int_{\sf D}\!\!\!\!d\xi\:
w(x,0;\xi)\, \Delta\! f\!(\xi\!,\!t)+\displaystyle\int_0^{t}\!\!\!\!d\tau \!\!\!\displaystyle\int_{\sf D}\!\!\!\!d\xi\:
w_t(x,t\!\!-\!\!\tau;\xi)\, \Delta\! f\!(\xi\!,\!\tau)\\[8pt]
& = & \displaystyle\int_0^{t}\!\!\!\!d\tau \!\!\!\displaystyle\int_{\sf D}\!\!\!\!d\xi\:
w_t(x,\! t\!\!-\!\!\tau;\! \xi)\, \Delta\! f\!(\xi\!,\!\tau)
\ea                                               \label{DeltaTv}
\eeq
Inequality (\ref{thetaineq}) implies for all $(x,t)\!\in\!D_T$ 
\ $\int_{\sf D} |w(x,t;\xi)| d\xi \le 2 N(t)$ \  
and, by (\ref{Deltaf}-\ref{DeltaTv}),
 \bea
&& \big| [{\cal T}\!v_1\!-\!{\cal T}\!v_2](x,t) \big|e^{-\lambda t}
\leq \displaystyle\int_0^{t}\!\!\!\!d\tau e^{-\lambda t}\!\!\!
\displaystyle\int_{\sf D}\!\!\!\!d\xi\:
|w(x,t\!-\!\tau;\xi)|\, \left| \Delta\! f\!(\xi\!,\!\tau)\right| \nn 
&&\qquad\qquad \leq 
% &\leq &  
\mu \|v_1\!-\!v_2\|_{\lambda, {\scriptscriptstyle T}} \!\!
 \int_0^t \!\!e^{-\lambda (t-\tau)}d\tau\!\!\displaystyle\int\limits_{\sf D}\!\! 
|w(x,t\!-\!\tau;\xi)|d\xi \nn 
&& \qquad\qquad \leq  \mu\|v_1\!-\!v_2\|_{\lambda, {\scriptscriptstyle T}}\!\! \int_0^t \!\!e^{-\lambda (t-\tau)} 2 N(t\!-\!\tau)  d\tau \nn
%&=& \mu\|v_1\!-\!v_2\|_{\lambda, {\scriptscriptstyle T}}\!\! \int_0^t \!\!e^{-\lambda \tau} 2 N(\tau)  d\tau 
&&\qquad\qquad \leq \frac{2\mu M'}{\lambda}\|v_1\!-\!v_2\|_{\lambda, {\scriptscriptstyle T}}, \qquad 
M':=M\!+\!\left\{\!\!\ba{ll} 0\quad &\mbox{DBC},\\
 a^{-1}\quad &\mbox{PBC, NBC and }a\neq 0,\\
 \lambda^{-1}\quad &\mbox{PBC, NBC and }a= 0.
\ea\right.   \qquad      \label{normDeltaTv}
\eea
On the other hand,  $\theta(0,i\eta)$ is a positive strictly decreasing function of $\eta>0$ 
fulfilling the property \ $\theta(0,\eta)\left(-i\eta\right)^{\frac 12}=\theta\!\left(\!\frac z\eta,\frac {-1}\eta\!\right)$, \ (see e.g. \cite{Mum83}, p. 33).  For \ $\tau\!\in\![0,T]$ it follows
$$
%\theta(0,\eta)\left(-i\eta\right)^{\frac 12}=\theta\!\left(\!\frac z\eta,\frac {-1}\eta\!\right)
%\qquad\Rightarrow\qquad 
\left(\frac{2\varepsilon \tau}\pi\right)^{\frac 12}\theta\!\left(\!0,i\frac 2\pi\varepsilon \tau\!\right)=\theta\!\left(\!0, \frac {i\pi}{2\varepsilon \tau}\!\right)
\le\theta\!\left(\!0, \frac {i\pi}{2\varepsilon T}\!\right)\!=:\! \frac 14
\left(\frac{\varepsilon}{2\pi}\right)^{\frac 12}\Theta^2;
$$
hence and
from  (\ref{thetax})$_2$ we find for \ $\tau\!\in\![0,T]$
\bea
2\pi\Vert \vartheta_{\tau}(\cdot,\tau) \Vert_2\!\le\! \sqrt{\kappa\!+\! 8e^{\frac {4\tau}\varepsilon}
\theta\!\left(\!0,i\frac 2\pi\varepsilon \tau\!\right)}
\!<\!\sqrt{\kappa}\!+\! e^{\frac {2\tau}\varepsilon}\sqrt{8\theta\!\left(\!0,i\frac 2\pi\varepsilon \tau\!\right)}\!\le\!
\sqrt{\kappa}\!+\! \Theta e^{\frac {2\tau}\varepsilon}\tau^{-\frac 14}.       \label{boundtht}
\eea
Using Schwarz inequality and (\ref{thetax}), (\ref{Greenf}), (\ref{Deltaf}), (\ref{boundtht})
eq.  (\ref{DeltaTv}) imply
\bea
& &\left|[{\cal T}\!v_1\!-\!{\cal T}\!v_2]_x(x,t) \right|e^{-\lambda t}
%\nn &&\qquad\qquad 
\leq\displaystyle\int_0^{t}\!\!\!\!d\tau \left|\displaystyle\int_{\sf D}\!\!\!\!d\xi\:
w_x(x,t\!-\!\tau;\xi)\, \Delta\! f\!(\xi\!,\!\tau) \right|e^{-\lambda t}\! \nn 
%&&\qquad\qquad 
&&\leq \displaystyle\int_0^{t}\!\!\!\!d\tau \, 2\pi\Vert \vartheta_x(\cdot,t\!\!-\!\!\tau)\Vert_2\:
\Vert\Delta\! f(\cdot,\tau)\Vert_2 e^{-\lambda t}  
%\displaystyle\int_0^{t}\!\!\!\!d\tau \, 2\pi\!\left\Vert \vartheta_x(\cdot,t\!-\!\tau)
%\right\Vert_2e^{-\lambda(t\!-\!\tau) } \,\mu\Vert v_1\!-\!v_2\Vert_{\lambda, {\scriptscriptstyle T}} 
\le \mu\Vert v_1\!-\!v_2\Vert_{\lambda, {\scriptscriptstyle T}} \,\left(\!2\!+\! \frac{12\!+\! 2\pi^2}{3\varepsilon}\!\right)^{\frac 12}
\!\!\displaystyle\int_0^{t}\!\!\!\!d\tau \,e^{-\lambda(t\!-\!\tau) } \,\nn &&
 \le  \frac{\mu}{\lambda}\left(\!2\!+\! \frac{12\!+\! 2\pi^2}{3\varepsilon}\!\right)^{\frac 12}
 \|v_1\!-\!v_2\|_{\lambda, {\scriptscriptstyle T}} \label{normDeltaTvx}\\
&&\Big|[{\cal T}\!v_1\!-\!{\cal T}\!v_2]_t(x,t)\Big|e^{-\lambda t}\le
\displaystyle\int_0^{t}\!\!\! d\tau \left| \displaystyle\int_{\sf D}\!\!\!\!d\xi\:
w_t(x,\! t\!\!-\!\!\tau;\! \xi)\, \Delta\! f\!(\xi\!,\!\tau)\right|e^{-\lambda t}
%\left|\sum\limits_{n\in\mathbb{Z}} \dot H_n\!(t\!\!-\!\!\tau)\,\Delta\! f_n\!(\tau)\,e^{in x}\right|\le
%\displaystyle\int_0^{t}\!\!\! d\tau \, \left[\sum\limits_{n\in\mathbb{Z}}\dot H_n^2\!(t\!\!-\!\!\tau)
%\right]^{\frac 12} \left[\sum\limits_{n\in\mathbb{Z}}\Delta\! f_n^2\!(\tau)\right]^{\frac 12}
\nn &&
\le\displaystyle\int_0^{t}\!\!\! d\tau \, 2\pi\Vert \vartheta_t(\cdot,t\!\!-\!\!\tau)\Vert_2\:
\Vert\Delta\! f(\cdot,\tau)\Vert_2 e^{-\lambda t}
%\le 2\pi\mu\Vert v_1\!-\!v_2\Vert_{\lambda, {\scriptscriptstyle T}} \displaystyle\int_0^{t}\!\!\! d\tau \, \Vert 
% \vartheta_t(\cdot,t\!\!-\!\!\tau)\Vert_2 \, e^{-\lambda(t\!-\!\tau) }
\le\mu\Vert v_1\!-\!v_2\Vert_{\lambda, {\scriptscriptstyle T}} \displaystyle\int_0^{t}\!\!\! d\tau \left(\!\sqrt{\kappa}\!+\! 
\Theta e^{\frac {2\tau}\varepsilon} \tau^{-\frac 14}\!\right) e^{-\lambda\tau }\nn &&
\le\mu\!\! \ba{l} \left[\!\frac{\sqrt{\kappa}}{\lambda}\!+\! \left(\! \lambda\!-\!\frac 2\varepsilon\!\right)^{-\frac 34}
\!\Theta\,\Gamma\!\left(\!\frac 34\!\right) \!\right] \ea    
 \Vert v_1\!-\!v_2\Vert_{\lambda, {\scriptscriptstyle T}}             \label{normDeltaTvt}
\eea 
[in the last step we have assumed $\varepsilon\lambda\!>\!2$ and used the gamma function 
$\Gamma(z)=\int^\infty_0\!dy\, e^{-y}y^{z-1}$]. Eq. (\ref{normDeltaTv}-\ref{normDeltaTvt}) imply
\bea
&&
\ba{l}\|{\cal T}v_1\!-\!{\cal T}v_2\|_{\lambda, {\scriptscriptstyle T}} \:  \leq \: \left\{\frac{\mu}{\lambda} 
\left[2 M'\!+\!\left(\!2\!+\! \frac{12\!+\! 2\pi^2}{3\varepsilon}\!\right)^{\frac 12}\!+\!
\sqrt{\kappa}\right]\!+\!\left(\! \lambda\!-\!\frac 2\varepsilon\!\right)^{-\frac 34}
\!\!\Theta\,\Gamma\!\left(\!\frac 34\!\right) \right\}\:   \|v_1\!-\!v_2\|_{\lambda, {\scriptscriptstyle T}}.  
\ea  \qquad     \label{217}
 \eea
Hence $\T$ is a contraction of $\B$ into itself provided we choose $\lambda\!>\!2/\varepsilon$ so large that
the coefficient of $\|v_1\!-\!v_2\|_{\lambda, {\scriptscriptstyle T}}$ at the rhs(\ref{217}) is smaller than 1.
Then, applying the fixed point theorem we find that
 there exists a unique solution
 of the problem $\T u=u$ in $\B$, i.e. of  (\ref{inteq'})
in the time interval $[0,T]$, for any
 $T>0$, and therefore in all $I=[0,\infty[$. \hfill \sq

\medskip
The existence and regularity of the solution {\it for all $t>0$} crucially depends
on the assumption that $f$ fulfills the Lipschitz condition (\ref{Lipsch}).
As known, if we assumed  $f$ to fulfill Lipschitz condition (\ref{Lipsch})
only {\it locally}\footnote{If for any bounded set 
$\Omega\subset D\times I\times \mathbb{R}$ 
there exists a constant $\mu$ {\it depending on $\Omega$} such that for any 
$(x,\!t,\!u_1), (x,\!t,\!u_2)\in\Omega$ (\ref{Lipsch})  is satisfied,
then  $f$ is said to satisfy a {\it local} Lipschitz condition w.r.t. $u$. Similarly if
$f$ depends on $x,t,u,u_x,u_{xx},u_t$.
}, then  in general the fixed
point theorem would be applicable only for a not too large $T$; 
as a consequence, one could not
exclude the occurrence of {\it blow-up's} \cite{Ali95}, i.e. singularities
of $u$ or its derivatives, for sufficiently
large $t$.

\end{document}